# Kinematic and volumetric analysis of coupled transmembrane fluxes of binary electrolyte solution components

Andriy E. Yaroshchuk [a,b], Stanislaw Koter [c], Volodymyr I. Kovalchuk [d*], and Emiliy K. Zholkovskiy [d]

[a] ICREA, 08010, Barcelona, Spain
[b] Department of Chemical Engineering, Universitat Politècnica de Catalunya, 08028, Barcelona, Spain
[c] Nicolaus Copernicus University, 87100 Toruń, Poland
[d] Institute of Biocolloid Chemistry of National Academy of Sciences of Ukraine, 03142 Kyiv, Ukraine

**Abstract**

The paper deals with relationships between the individual transmembrane fluxes of binary electrolyte solution components and the experimentally measurable quantities describing rates of transfer processes, namely, the electric current, the transmembrane volume flow and the rates of concentration changes in the solutions adjacent to the membrane. Also, we collected and rigorously defined the kinetic coefficients describing the membrane selective and electrokinetic properties. A set of useful relationships between these coefficients is derived.

An important specificity of the proposed analysis is that it does not use the Irreversible Thermodynamic approach by analyzing no thermodynamic forces that generate the fluxes under consideration. Instead, all the regularities are derived on the basis of conservation and linearity reasons. The terminology "Kinematics of Fluxes" is proposed for such an analysis on the basis of the analogy with Mechanics where Kinematics deals with regularities of motion by considering no mechanic forces. The only thermodynamic steps of the analysis relate to the discussion on the partial molar volumes of electrolyte and ions that are the equilibrium thermodynamic parameters of the adjacent solutions. These parameters are important for interrelating the transmembrane fluxes of the solution components and the transmembrane volume flow. The paper contains short literature reviews concerned with the partial molar volumes of electrolyte and ions: the methods of measurement, the obtained results and their theoretical interpretations. It is concluded from the reviews that the classical theories should be corrected to make them applicable for sufficiently concentrated solutions, 1M or higher. The proposed correction is taken into account in the kinematic analysis.

**Content**

## 1. Introduction: Irreversible Thermodynamics and Kinematic of Fluxes

*1.1 Membrane as electro- mechano- chemical transducer*

Membranes are the layers of diverse nature that allow components of mixtures to be transported through such layers in other proportions than those defined by the properties of components in the mixture. This property motivated the use of such selective membranes in a variety of technologies dealing with separation of components of various mixtures without the changes of aggregate state. In particular, the components of electrolyte solutions are often separated by using electrically- and pressure- driven membrane processes [1-3].

In many technological processes, membrane separation of mixtures is conducted by consuming the external thermodynamic work which is partially stored in the system in the form of additional Free Energy of separation products with reference to the initial mixture. Moreover, a system of two solutions with different compositions always bears the additional Free Energy with reference to their mixture. Consequently, by properly mixing these solutions, the system can produce electrical or mechanical work. Such reversible (completely or partially) processes can be organized with help of membranes having special selective properties with respect of the solutions being mixed.

The abovementioned thermodynamic opportunity to obtain electrical and mechanical work by mixing solutions is widely employed in analytical technique dealing with, respectively, the ion-selective membrane electrodes [4] and the membrane osmometers for studying macromolecular solutes [5]. Obviously, in the latter applications, the additional free energy is converted to the work in low amounts that are insufficient for using in the Energy Industry. However, in $1950^{th}$ and $1970^{th}$, it was suggested to utilize the giant Free Energy accumulated in the sea water by reversible mixing it with the water from other natural sources having much lower salinity [6,7]. It was proposed to conduct such a membrane mixing with obtaining electrical [6] or mechanical [8] work. In $2000^{th}$, the interest to this project increased, and, during the past decades, it became in the focus of intensive studies [9-15].

The above statement demonstrates the ability of membranes to serve as an electrochemical and/or mechano-chemical transducer. While using membranes in contact with two electrolyte solutions, one often observes the Electrokinetic Phenomena (Electroosmosis, Streaming Potential and Current etc.) [16,17]. Hence, such a membrane is also an electro-mechanical transducer.

The behavior of membranes as electro-mechano-chemical transducer was discussed above for the synthetic membranes employed in different technologies. It should be noted that a similar coupling between the electrical, mechanical and chemical processes exists in the case of biological membranes that play an important role in the living activity of biological cells [18]. Remarkably, the membranes discussed above are very different by their chemical origin, morphology and physical properties, but they show similar behavior in terms of Thermodynamics. Therefore, more than six decades ago, Thermodynamics was chosen to be the theoretical basis for studying Membrane Phenomena (MP). Since all the MP are observed under non-equilibrium conditions, the Linear Irreversible Thermodynamics (LIT) [19,20] was considered as a proper tool for addressing Membrane Phenomena.

*1.2 Linear Irreversible Thermodynamics for addressing membrane phenomena*

Within the frameworks of the LIT, the membrane transport of solution components is described with the help of $n \times n$ matrix, $\mathbf{L}$, whose elements are referred to as the kinetic coefficients. This matrix is represented in the linear relationship

$$\mathbf{Y^T} = \mathbf{L}\mathbf{X^T} \tag{1}$$

where the column vector, $\mathbf{Y^T}$, is a transposition of the row vector $\mathbf{Y} = \{Y_1, Y_2, ... Y_n\}$ which yields a set of Thermodynamic Fluxes, describing the rates of the solution component transfer through the membrane. The column vector, $\mathbf{X^T}$ is the transposed raw vector $\mathbf{X} = \{X_1, X_2, ... X_n\}$ which yields a set of Thermodynamic Forces that are the differences between some intensive thermodynamic parameters, $\Omega_k$, attributed to the solutions adjacent to the membrane, $X_k = \Delta\Omega_k = \Omega'_k - \Omega''_k$, (Fig.1). Hereafter, the notations (′) and (″) signify, respectively, the quantities attributed to the agreed left and right hand solutions adjacent to the membrane (Fig.1).

The vector of Thermodynamic Forces, $\mathbf{X}$, describes the external influences resulting in the system departure from the thermodynamic equilibrium state. Accordingly, at $\mathbf{X} = 0$, the system remains in the thermodynamic equilibrium state when all the fluxes are absent, i.e., $\mathbf{Y} = 0$. Thus, the right hand side of Eq.(1) yields the linear terms in the Taylor series expansion of Thermodynamic Fluxes in terms of Thermodynamic Forces. Accordingly, while using the linear relationship given by Eq.(1), the departure from the thermodynamic equilibrium is assumed to be small, and the matrix of kinetic coefficients is considered to be independent of external driving forces. Therefore, $\mathbf{L}$ is completely defined by the membrane properties. It should be added that in many cases, accounting for the linear terms only is insufficient for describing the system behavior.

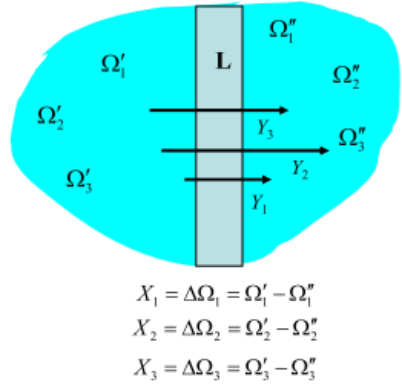

$$X_1 = \Delta\Omega_1 = \Omega'_1 - \Omega''_1$$
$$X_2 = \Delta\Omega_2 = \Omega'_2 - \Omega''_2$$
$$X_3 = \Delta\Omega_3 = \Omega'_3 - \Omega''_3$$

**Fig.1. Thermodynamic Fluxes and Forces**

By using the measured matrix of kinetic coefficients, $\mathbf{L}$, one can address all the membrane phenomena without knowledge about the membrane morphology and specific micro-mechanisms responsible for the membrane selectivity and permeance. At the same time, studies of such mechanisms can be conducted separately by analyzing the influence of membrane morphology and the external conditions on the matrix of kinetic coefficients, $\mathbf{L}$. In the literature, there are hundreds of publications concerned with predictions of kinetic coefficients on the basis of a variety of assumed models of the membrane morphology. As a simple example, we will just mention only two of them where a membrane is considered to be a packed bed of charged solid spheres [21,22].

A substantial progress in the Membrane Science was achieved due to the use of an approach developed by Onsager who suggested a method of choosing convenient sets of Thermodynamic Fluxes and Forces, $\mathbf{Y}$ and $\mathbf{X}$ [23, 24].

In refs.[23,24], Onsager considered the entropy, $s$, changes in a thermodynamically open system involved in irreversible processes. These changes can be subdivided in two types: (i) those produced due to the heat and mass exchange with a thermostat, $ds_{ex}$, and (ii) those generated inside the system, $ds_{int}$. Hereafter, to the rate of entropy changes of the second type, $ds_{int}/d\tau$ (where $\tau$ is the time), we will refer as the Entropy Production Function (EPF), $W = ds_{int}/d\tau$. For the LIT case, the function, $W$, is a positive definite bilinear form of Thermodynamic Fluxes and Forces.

Importantly, in some special cases, such a bilinear form takes the diagonal form
$$W = \mathbf{X} \cdot \mathbf{Y} \tag{2}$$
According to the Onsager theorem [23,24], if, and only if, the EPF takes the form given by Eq.(2), the matrix $\mathbf{L}$ has the diagonal symmetry, $\mathbf{L} = \mathbf{L}^T$.

Thus, while conducting the LIT analysis, it is convenient to choose the sets of Thermodynamic Forces and Fluxes in the manner allowing representing the EPF in the diagonal form of Eq.(1). Such a choice enables one to reduce the number of independent kinetic coefficients substantially, by using the abovementioned symmetry rule, $\mathbf{L} = \mathbf{L}^T$. Each of the independent coefficients $L_{nk}$ can be measured with the help of an experimental scheme defined by Eq.(1). While using the symmetry rule, obtaining a cross coefficient, $L_{nk}$ ($n \neq k$), simultaneously gives the value of the respective symmetric coefficient, $L_{kn}$.

Starting with the middle of previous century, the LIT approach based on the Onsager theorem [23,24] has been widely used for analyzing the MP for a membrane placed between two electrolyte solutions. In the studies of Mazur & Overbeek [25] and Lorenz [26], by using this theorem, the authors interrelated the kinetic coefficients describing different Electrokinetic Phenomena [27, 28]. It was demonstrated that the Electroosmotic and Streaming Potential Coefficients are equal regardless the membrane origin and morphology. Also, the authors interrelated the membrane electric conductances measured at zero transmembrane volume flow and pressure difference as well as the hydraulic permeances measured at zero transmembrane electric current and voltage.

In aforementioned refs.[25,26], the LIT analysis was conducted for the case when a membrane is placed between two solutions having the same compositions as it is typical for Electrokinetics [16, 17]. A more general case was addressed in the pioneering paper of Staverman [27]. He used the LIT approach for analyzing the transport of components of a mixed electrolyte solution driven through the membrane by the electric potential, pressure and ion concentration differences. While considering the limiting case of equal compositions of the adjacent electrolyte solutions, he rederived the results obtained for the Electrokinetic Phenomena in refs.[25, 26].

The starting point of Staverman theory was the expression for the EPF having the diagonal bilinear form given by Eq.(2) where the vectors of the Thermodynamic Forces and Fluxes were given as

$$\mathbf{X} = \{\Delta \mu_k\}$$
$$\mathbf{Y} = \{J_k\}$$
(3)

where $J_k$ and $\Delta \mu_k$ are, respectively, the kth solution component transmembrane flux and chemical potential difference across the membrane. By using such a form for the EPF, Staverman [27] derived very general and elegant thermodynamic expressions for addressing different MP.

Later, by transforming the EPF given by Eqs.(2) and (3), Kedem & Katchalsky and Michaeli & Kedem in their famous publications [28-31] derived several equivalent versions of LIT equation sets describing the MP for membrane between two binary mixtures of the same non-electrolyte substances [28 29] and between two solutions of the same binary electrolyte [30,31]. Hereafter, we will refer to these studies and results as the KKM theory.

Detailed discussion of the KKM results is given in Section 3. Now, we only mention that the Thermodynamic Fluxes were obtained in [30, 31] to be linear combinations of three solution component fluxes, $J_k$: two ions and solvent. The obtained combinations are interpreted by the authors as the transmembrane solution volume flow, the electrolyte (salt) flux and electric current. As demonstrated later [32-34], strictly speaking, the first and second combinations have the abovementioned physical meanings at zero current, only. However, when electric current is not zero, the interpretation suggested in [30, 31] can serve as a certain approximation which allows the KKM fluxes and forces to be linked to a strictly defined thought experiments.

Within its validity range the KKM theory gives scheme of imposing and/or measuring the Forces and Fluxes. The latter enabled the authors of [29-31] to introduce a set of six coefficients, namely, *the reflection coefficient; the hydraulic and osmotic permeances; the electric conductance, the electric transport number for any of two ions*; *the Electroosmotic Coefficient*. The introduced coefficients have clear physical meaning and are referred to as *the practical coefficients* in the relevant literature. Due to the symmetry of the obtained $3 \times 3$ matrix, **L**, the six practical coefficients are sufficient to express all the nine matrix elements.

The wide field of applicability and the fact that the KKM equations contain only the directly measurable and strictly defined quantities explain the great impact produced by the KKM theory on the development of Membrane Science. The KKM equations triggered a powerful flow of publications concerned with different aspects of Membrane Science and defined its development for decades. References [35-46] comprise papers on membranes having different nature and represent a small portion of the tremendous massive of studies using the KKM theory that have been reported during the past two decades, only.

*1.3 Kinematics of Fluxes and Membrane Phenomena*

In different cases, for addressing electro-mechano-chemical transduction discussed in Section 1.1, there is no need to use the scheme based on the Onsager theorem and outlined in the previous Section. Some of the Membrane Phenomena manifest themselves as coupling between transmembrane fluxes of different physical nature. In this paper, we use the terminology "Kinematics of Fluxes" for signifying an analysis of membrane phenomena in terms of transmembrane fluxes only, i.e., by considering no thermodynamic forces that generate these fluxes. Such a terminology is based on the analogy with Mechanics where Kinematics deals with regularities of motion by considering no mechanic forces defining the regularities.

As stated in Section 1.1, the ability of membranes to transform free energy from one form to another exists due to the membrane property to provide transfer of the adjacent solution components in other proportions than those defined by the properties of the solutions. These proportions can be characterized by a set of coefficients that show the contributions of each of the solution component flux into a measurable transmembrane flux. Accordingly, these coefficients bear information about selective properties of membranes with respect of the transported solution components.

Two of three measurable Thermodynamic Fluxes of the KKM theory [30,31], namely, the electric current and the volume flow can be imposed and controlled by means of external electric and hydraulic sources. Below, we present a list of membrane phenomena to be analyzed within the frameworks of Kinematic of Fluxes. These phenomena are observed while imposing electric current or volume flow:

(a) *Streaming Current* is the transmembrane electric current driven by the volume flow passed through the membrane;

(b) *Electroosmosis* is the transmembrane volume flow which is driven by the electric current passed through the membrane;

(c) *Reverse Osmosis* is the composition changes produced in the solutions adjacent to a membrane when the transmembrane solution volume flow is imposed at zero transmembrane electric current;

(d) *Electrodialysis Effect* is a change of composition which is produced by the transmembrane electric current in the solutions adjacent to membrane.

For addressing the effects listed above, one can introduce a set of coefficients characterizing a given membrane. The coefficients should be introduced with the help of certain thought experiments with the membrane. As well, it is possible to introduce a set of relationships between these coefficients while taking into account that, in the linear case, one deals with a superposition of the effects listed above.

The scheme presented in Fig.2 illustrates the complex couplings that can be observed between the transmembrane electric current and volume flow and lead to combining the above listed effects. For example, passing the electric current through a membrane under certain conditions gives rise to both the volume flow and concentration changes due to the Electroosmosis and Electrodialysis, respectively. Simultaneously, the produced volume flow additionally contributes to the concentration changes due to Reverse Osmosis. Thus, one can expect that the membrane parameters responsible for Electroosmosis, Electrodialysis and Reverse Osmosis can be interrelated with each other.

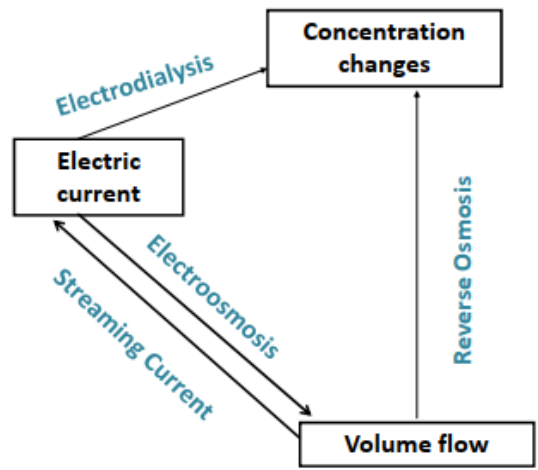

**Fig. 2 Electro-mechano-chemical effects due to the complex coupling between transmembrane electric current and volume flow**

Thus, it is important to establish relationships between the abovementioned coefficients that, on the one hand, describe the membrane selective properties and, on the other hand, reflect the membrane ability to conduct mechano-electro-chemical transformations of free energy. One might add that, once obtaining such relationships with the help of the Flux Kinematic analysis, one can use them in the LIT analysis based on the Onsager cross relationships.

The Flux Kinematic analysis requires knowing strict relationships between the individual component transmembrane fluxes and the measured and imposed fluxes. For effects (a)-(d) listed above, the latter fluxes are the transmembrane electric current and volume flow. As stated in Section 1.2 by referencing to [32-34], for non-zero transmembrane electric current, the KKM theory gives only an approximate expression for the transmembrane volume flow in terms of the solution component fluxes. Therefore, it is required to derive a more general expression than that given by the KKM theory and analyze how important is the correction to be obtained. It should also be noted that the transmembrane volume flow gives the rate of the volume changes of the solutions adjacent to the membrane. These rates are equal by magnitude and opposite by sign due to the assumed incompressibility of the solutions [47]. At sufficiently high electrolyte concentrations, it is expected that adding/removing electrolyte into/from a solution affects the volume of solution, noticeably.

While considering the transmembrane volume flow, for analyzing the adjacent solution volume changes, it is correct to use the Reversible Thermodynamics (Thermostatics) of the electrolyte dissolving. The latter is true since the LIT analysis of purely membrane transport is always conducted by assuming the adjacent solutions to be in the equilibrium states that are close to each other. Accordingly, one can use for such an analysis all the nontrivial results that have been reported in the literature during over the century [48-78]. The results obtained in these studies reveal the importance of the effects originating from the reorganization of solvent by the dissolved electrolyte, in particular, by electrostriction produced by the individual ions. The latter effect was predicted in the pioneering paper of Nernst & Drude [48].

Thus, it is important to obtain a correct expression for transmembrane volume flow with accounting of the results of studies in Thermostatics of electrolyte solutions [48-78] and evaluate the applicability limits of the KKM theory [30, 31] on this basis.

In the case of non-zero electric current, a difficulty in the abovementioned Thermostatic analysis of adjacent solutions is associated with unavoidable contributions of the source and sinks of electric current into the changes of the solution ionic compositions. These changes occur additionally to those resulting from the membrane transport of ions and are defined by the proportion according to which the ions transfer the current through the abovementioned source and sink.

In order to avoid the problem outlined above, the authors of the KKM theory considered the membrane with the adjacent solutions inside an electrochemical cell made up by a couple identical electrodes being the said source and sink. The electrodes are assumed to be reversible for one and indifferent for other of the binary electrolyte ions. Essentially, the KKM theory gives the LIT description of such a cell with a membrane, not the membrane itself. The abovementioned discrepancy between the KKM and the transmembrane volume flows is one of examples of that. In Section 3, we will give a detailed critical analysis of fluxes considered in [30, 31] as the transmembrane ones.

The electrochemical electrode couple introduced in the KKM theory can be understood as an element of a real experimental set up. At the same time, the electrochemical cell made up by such electrodes can be interpreted as a purely hypothetical construct. Accordingly, introducing such a cell in the KKM theory can be considered as a convenient theoretical step for analyzing regimes with non-zero current. Therefore, we will call such a hypothetical cell, as the Virtual Electrochemical Cell (VEC). For addressing the purely membrane transport with the help of VEC, the relevant theoretical analysis should include the following steps:
- to assume certain properties of the VEC electrodes;
- to interrelate the transmembrane fluxes and the rates of composition changes of the adjacent solutions;

- to extract the regularities that are inherent in the purely membrane transport from the information contained by the solution composition changes;
- to ensure that the obtained regularities of membrane transport are independent of the assumed properties of VEC.

The reasons stated in the presented section motivated us to use the purely kinematic approach for conducting a systematical analysis of several important aspects of membrane transport. Next, we will state these aspects more specifically.

*1.4 Objectives and structure of the paper*

The present study is intended to achieve following objectives.

1) To interrelate the individual transmembrane fluxes of the solution components with the measurable or imposed rates that are referred to as the transmembrane electric current, solution volume flow and electrolyte flux. For the electric current, such an interrelation is trivial, but, for the other two rates, the results presented in literature require to be extended for highly concentrated solution.

2) To collect the membrane parameters that are responsible for the selectivity of transport of the individual solution components in the regimes of the imposed transmembrane electric current and volume flow.

3) To address the coupling between the transmembrane electric current, solution volume follow and electrolyte flux, in terms of the coefficients responsible for the membrane selectivity with respect to different solution components. Such an analysis will give the description of electro-mechano- chemical membrane phenomena listed in Section 1.3, as (a)-(d).

4) By using the purely kinematic analysis, i.e., without the use of the Onsager theorem, to establish relationships between different parameters responsible for the membrane selectivity as well as between these parameters and the coefficients describing the electro-mechano-chemical coupling of fluxes.

The objectives formulated above define the communication structure.

In Section 2, we introduce the VEC, which was mentioned in the end of Section 1.3, and make some basic definition related to the rates of ion and electrolyte concentration changes in the VEC and individual ionic fluxes.

In Section 3, the rate of volume changes in the compartments of the VEC is discussed. The discussion employs an equilibrium thermodynamic parameter which is attributed to the adjacent solutions and is referred to as the partial molar volume of electrolyte. In our opinion, the latter quantity deserves more attention than it is usually paid in the membrane literature. Therefore,

Section 3 contains a short survey of the literature concerned with the theoretical basis for measuring the electrolyte partial molar volume, the respective experimental data and explanations of their behavior. As well, Section 3 contains a detailed discussion of the KKM theory. The discussion leads to conclusions regarding both the advantages and the restrictions of the KKM theory.

One of the conclusions formulated in Section 3 is that the KKM theory requires an extension to be capable of addressing sufficiently concentrated electrolyte solutions. In Section 4, it is pointed out that, to this end, one should unavoidably use such parameters as the partial molar volumes of individual ions. Accordingly, in Section 4, a survey of the literature is presented about the nontrivial approaches to the experimental determining these parameters and a summary of results reported in the literature for different ions. The survey also includes theoretical models intended to describe different unexpected experimental results. Transmembrane volume flow, which is written in Section 4 in terms of the partial molar volumes of individual ions, is evaluated by using the literature results on the volumes.

In the Section 5, we introduce sets of coefficients that describe selective properties of membrane with reference to the solution components and describe in these terms mechano-chemical, electrochemical and electro-mechanical coupling between the fluxes. As well, a set of important relationships is derived to interrelate the coefficients describing the membrane selectivity under various imposed conditions.

## 2. Fluxes of ions and electrolyte

We consider a thought experiment employing two compartment cell sketched in Fig.3. The compartments are separated from each other by the membrane under consideration and filled by binary solutions of the same strong binary electrolyte, $A_{\nu_A} B_{\nu_B}$, dissociating according to the stoichiometric equation

$$A_{\nu_A} B_{\nu_B} \rightarrow \nu_A A + \nu_B B \tag{4}$$

where the integer numbers $\nu_A > 0$ and $\nu_B > 0$ are the stoichiometric coefficients. The electroneutrality of initial electrolyte implies the validity of the relationship

$$\nu_A z_A + \nu_B z_B = 0 \tag{5}$$

where $z_A$ and $z_B$ are the charge numbers of ions. For imposing the electric current and the volume flow through the membrane, the compartments are supplied by electrode and capillary couples, respectively.

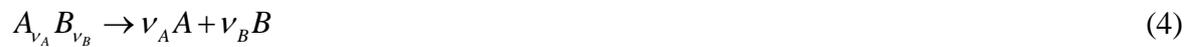

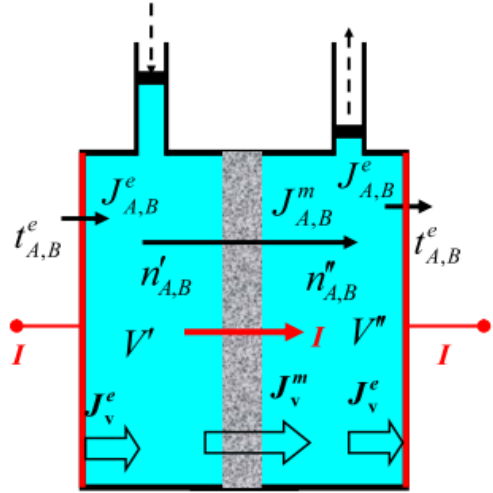

**Fig.3 Two compartment cell containing membrane**

In order to analyze kinetics of the pure membrane transport of the solution components, the rate of transport is assumed to be sufficiently slow to allow both the thermodynamic equilibrium and the local electroneutrality to be maintained within the compartments. Consequently, the solution components are assumed to be uniformly distributed within each of the solutions filling the compartments.

By using the stoichiometric and electroneutrality equations, Eqs.(4) and (5), respectively, and assuming the total electroneutrality of each of the solutions, one can interrelate the total amounts, $n_{A,B}(mol)$, of ions, $A$ or $B$, in any of the compartments and their spatially uniform molar concentrations, $c_{A,B}\left(mol/m^3\right)$

$$\frac{n_A}{\nu_A} = \frac{n_B}{\nu_B} = n \quad \text{(a)}$$

$$\frac{c_A}{\nu_A} = \frac{c_B}{\nu_B} = c = \frac{n}{V} \quad \text{(b)} \qquad (6)$$

$$c_{A,B} = \frac{n_{A,B}}{V} \quad \text{(c)}$$

where $V$ and $n$ are the volume of any of the compartments shown in Fig.3 and the electrolyte amount inside this volume, respectively.

## 2.1 Virtual Electrochemical Cell (VEC)

Our attention is focused on the solution component transport through the membrane. This transport gives rise to changes of the solution volume and component amounts in the adjacent solutions (Fig.3). In the presence of electric current, unavoidably, all these quantities are also changed due to the reactions at the electrodes of the VEC, which was already mentioned in Section 1.3, and now is used in the thought experiment sketched in Fig.3. We will refer to the total rates of these combined changes as the respective apparent fluxes. As stated in Section 1.3, within the frameworks of a theoretical scheme intended for obtaining transmembrane fluxes, introducing the VEC with given electrode properties can be considered as a purely theoretical step.

Consider the transmembrane fluxes of ions A and B, $J_A^m$ and $J_B^m$. These fluxes define the electric current, $I$, transferred through the membrane according to equation

$$I = F\left(J_A^m z_A + J_B^m z_B\right) \tag{7}$$

$F$ is the Faraday constant.

Since the electroneutrality of solutions is maintained, the magnitude of electric current remains the same in all crosssections of the cell sketched in Fig.3. However, through different crosssections, the current may be transferred by ions in different proportion than that given by Eq.(7).

The VEC electrodes are assumed to impose given portions of current being transferred through them by each of the ions, $t_A^e$ and $t_B^e$. It is also assumed that the solvent does not take part in the electrode reactions and, thus, $t_A^e + t_B^e = 1$. In particular, at the electrodes, individual ion fluxes are defined by the abovementioned electrode transport numbers $t_A^e$ and $t_B^e$ (Fig.3).

$$J_{A,B}^e = \frac{I t_{A,B}^e}{F z_{A,B}} \tag{8}$$

For simplicity, we consider the VEC electrodes having redox properties with respect to only one type of ion - either A or B. According to such an assumption, each of the fluxes at the electrode should always become zero at $I = 0$. Such a situation can take place for two types of the VEC's, only

$t_A^e = 0;\ t_B^e = 1$  \qquad VEC (A)

or  \hfill (9)

$t_A^e = 1;\ t_B^e = 0$  \qquad VEC (B)

Note that the cell notation proposed above is linked to the notation of the blocked ion, i.e., to the ion for which the electrode transport number is zero. Importantly, the VEC's defined by Eq.(9) have been employed in the KKM theory [28,31] in the same role. In the analysis below, we will use the general notation, $t_A^e$ and $t_B^e$, with understanding that only two abovementioned couples could be substituted for $t_A^e$ and $t_B^e$.

## 2.2. Transmembrane and apparent fluxes

Let us now combine Eq.(8) with the conservation laws written for each of the components of solutions inside the cell shown in Fig.3. Accordingly, the transmembrane fluxes of $A$ and $B$ ions, $J_A^m$, $J_B^m$ ($J_{A,B}^m$), and solvent, $J_w^m$, can be interrelated with the rates of changes of the ion and solvent amounts within the respective compartments of VEC, as

$$J_{A,B}^m = \frac{It_{A,B}^e}{Fz_{A,B}} + \frac{dn''_{A,B}}{d\tau} = \frac{It_{A,B}^e}{Fz_{A,B}} - \frac{dn'_{A,B}}{d\tau} \quad \text{(a)}$$

$$J_w^m = \frac{dn''_w}{d\tau} = -\frac{dn'_w}{d\tau} \quad \text{(b)} \tag{10}$$

where $\tau$ is the time and $n_w (mol)$ is the solvent amount in the respective compartment. The notations (′) and (″) signify the quantities attributed to the left and right hand side compartments shown in Fig.3, respectively.

By combining Eqs (6a) and (10a), one obtains:

$$\frac{dn''}{d\tau} = -\frac{dn'}{d\tau} = J^* \tag{11}$$

where we introduced the apparent electrolyte flux, $J^*$, which gives the rate of changing the electrolyte amounts, $n'$ and $n''$, in the VEC compartments (Fig. 3).

When the current, $I$, the electrode transport numbers, $t_{A,B}^e$, and the apparent electrolyte flux, $J^*$, are known, the actual transmembrane ion fluxes, $J_A^m$ and $J_B^m$, are reconstructed with the help of Eqs.(10a) and (11).

Using Eqs.(6a), (10a) and (11) leads to the following expressions for $J^*$

$$J^* = \frac{1}{2}\left(\frac{J_A^m}{\nu_A} + \frac{J_B^m}{\nu_B}\right) - \frac{I}{2F}\frac{t_A^e - t_B^e}{z_A \nu_A} = \frac{J_A^m}{\nu_A}t_B^e + \frac{J_B^m}{\nu_B}t_A^e \tag{12}$$

Inspecting two versions of the VEC given by Eq.(9) and using Eq.(12) give

$$J^* = J_A^m / \nu_A \qquad \text{VEC (A)}$$

or $\tag{13}$

$$J^* = J_B^m / \nu_B \qquad \text{VEC (B)}$$

As it is clear from the first equality in Eq.(12), at zero current, $I = 0$, the apparent electrolyte flux takes the form

$$\left(J^*\right)_{I=0} = \frac{J_A^m}{\nu_A} = \frac{J_B^m}{\nu_B} = J^m \tag{14}$$

where we introduced the actual transmembrane electrolyte flux, $J^m$, which has a physical meaning in the only case of zero current, $I = 0$, i.e., when the ions are transferred through membrane in stoichiometric amounts set by Eq.(4)

By using Eqs.(10)-(12), one can also obtain expressions for apparent fluxes of ions, $J_A^*$ and $J_B^*$

$$J_{A,B}^* = -\frac{dn'_{A,B}}{d\tau} = \frac{dn''_{A,B}}{d\tau} = J_{A,B}^m - \frac{It_{A,B}^e}{Fz_{A,B}} = J^* v_{A,B} \qquad (15)$$

## 3. Apparent volume flow

Simultaneously with the solute, the solvent is transferred through the membrane. However, for practical needs, it is often more convenient to deal with the transmembrane solution volume flow, $J_v^m$, rather than the sole solvent flux, $J_w^m$.

Recall that each of the adjacent solutions is assumed to be infinitely close to its equilibrium state. Hence, with the help of equilibrium equation of state, the solution volume can be expressed as a function of absolute temperature, *T*, pressure, *p*, and amounts of solvent and electrolyte, $n_w$ and *n*, respectively.

$$V = V(T, p, n, n_w) \qquad (16)$$

Consequently, a small change of the volume of any of the solutions can be represented as

$$dV = \left(\frac{\partial V}{\partial T}\right)_{p,n,n_w} dT + \left(\frac{\partial V}{\partial p}\right)_{T,n,n_w} dp + \left(\frac{\partial V}{\partial n}\right)_{T,p,n_w} dn + \left(\frac{\partial V}{\partial n_w}\right)_{T,p,n} dn_w \qquad (17)$$

We confine ourselves by analyzing the isothermal regimes that allows us to omit the first term on the right hand side of Eq.(17). As well, the compressibility of electrolyte solutions is close to that of water and takes value of order of $10^{-10} Pa^{-1}$ [47]. It means that even a high pressure difference about 100 *bar* can change the volume of a given solution by 0.1%, not more. Consequently, for lower pressure differences, one can omit the second term on the right hand side of Eq.(17). Next, while taking into account Eqs.(10b) and (11), one can obtain from Eq.(17) the following equality

$$\frac{dV''}{d\tau} = -\frac{dV'}{d\tau} = J_v^* \qquad (18)$$

where we introduced the apparent volume flow, $J_v^*$, describing the equal changes of solution volumes in the VEC compartments shown in Fig.3.

By inspecting the sketch in Fig.3, we present the following balance equation

$$J_v^m = J_v^* + J_v^e \qquad (19)$$

where $J_v^m$ is the transmembrane volume flow and $J_v^e$ is the volume flow associated with the electrode reactions.

Below in this Section, we consider how the apparent volume flow, $J_v^*$, is interrelated with the solvent and electrolyte fluxes.

### 3.1 Partial and apparent molar volume of electrolyte

Let us write Eq.(17) with omitted two first terms on the right hand side for each of two compartments shown in Fig.3. By combining the obtained equations written in terms of time derivatives with Eqs.(10b) and (17), we express the apparent volume flow, $J_v^*$, in the form

$$J_v^* = \mathrm{v} J^* + \mathrm{v}_w J_w^m \tag{20}$$

In Eq.(20), we introduced the partial molar volumes of electrolyte and solvent, $\mathrm{v}\,(m^3/mol)$ $\mathrm{v}_w\,(m^3/mol)$, defined, respectively, as

$$\begin{aligned}\mathrm{v} &= \left(\frac{\partial V}{\partial n}\right)_{T,p,n_w} &\text{(a)}\\ \mathrm{v}_w &= \left(\frac{\partial V}{\partial n_w}\right)_{T,p,n} &\text{(b)}\end{aligned} \tag{21}$$

At constant temperature and pressure, $V = V(n, n_w)$. It can be shown that $V(n, n_w)$ is the first order Euler homogeneous function, $V(kn, kn_w) = kV(n, n_w)$. Such a property implies that

$$V(n, n_w) = \left(\frac{\partial V}{\partial n}\right)_{T,p,n_w} n + \left(\frac{\partial V}{\partial n_w}\right)_{T,p,n} n_w \tag{22}$$

By combining Eqs.(6b), (21) and (22), one obtains a useful relationship

$$c\mathrm{v} + c_w \mathrm{v}_w = 1 \tag{23}$$

where $c_w = n_w / V\,(mol/m^3)$ is the solvent concentration. While using Eqs.(20) and (23), we express the transmembrane solvent flux through the apparent volume and electrolyte fluxes, as

$$J_w^m = \frac{J_v^* - \mathrm{v} J^*}{1 - c\mathrm{v}} c_w \tag{24}$$

As it follows from Eq.(24), the electrolyte partial molar volume, v, is an important quantity enabling one to interrelate the directly measurable apparent fluxes, $J_v^*$ and $J^*$, and the transmembrane solvent flux, $J_w^m$. Clearly, the role of this quantity becomes more important with increasing the electrolyte concentration, $c$.

As we mentioned in Section 1.3, the partial molar volumes of different electrolytes have been intensively studied during a long period of time [48]-[78]. One of the nontrivial effects discovered in these studies was that the electrolyte partial molar volume was found to be a function of electrolyte concentration, $v = v(c)$. This function is obtained from the concentration dependency of the apparent molar (molal, in many publications) volume, $v^* = v^*(c)$, which is the solution volume, $V$, change compared to the pure solvent volume, $V_w$, calculated per 1 mole of the solute amount. The function $v^*(c)$ is determined from the measured mass density of electrolyte solution, $d(kg/m^3)$, and pure solvent, $d_w(kg/m^3)$, by using one of the expressions represented in the following chain of identities

$$v^*(c) = \left(\frac{V - V_w}{n}\right)_{p,T,n_w} = \frac{M}{d_w} - \frac{d(c) - d_w}{cd_w} = \frac{M}{d} - \frac{d(m) - d_w}{mdd_w} = v^*(m) \qquad (25)$$

where $M(kg/mol)$ is the solute molar mass; $m(mol/kg)$ is the solution molality. Detailed derivation of Eq.(25) is given in Appendix A1.

While realizing that $m = c/d_w(1 - cv^*)$ and using the Eqs.(23) and (25), one can interrelate the specific and apparent molar volumes, as

$$v = v^* + n\left(\frac{\partial v^*}{\partial n}\right)_{p,T,n_w} = v^* + m\left(\frac{\partial v^*}{\partial m}\right)_{p,T,n_w} = v^* + \frac{1 - cv^*}{1 + c^2\left(\frac{\partial v^*}{\partial c}\right)_{p,T,n_w}} c\left(\frac{\partial v^*}{\partial c}\right)_{p,T,n_w} \qquad (26)$$

The detailed derivation of Eq.(26) is also reproduced in the Appendix A1.

As it follows from Eq.(26), the partial and apparent molar volumes, $v^*$ and $v$, coincide when the measured value of $v^*$ is independent of the solute molar concentration $c$. Otherwise, they coincide in the limiting case of infinite dilution, $c \to 0$, only.

For the first time, the systematic experimental studies of apparent molar volume, $v_*$, based on the density measurements followed by the use of Eq.(26) have been reported by Mason [49]. For solutions of different electrolytes, by fitting the measured concentration dependencies, $v^*(c)$, Mason established the limiting law for high dilutions

$$\begin{aligned} v^* &= v(0) + K\sqrt{c} \qquad (a) \\ v &= v(0) + \frac{3}{2}K\sqrt{c} \qquad (b) \end{aligned} \qquad (27)$$

where $K(m^{9/2}/mol^{3/2})$ is an adjustable parameter. Consequently, the common value of the apparent and partial molar volumes at infinite dilutions, $v^*(0) = v(0)$, is obtained by extrapolating the measured dependency $v^*(\sqrt{c})$ to $c \to 0$. Note that Eq.(27b) can be obtained by substituting (Eq.27a) into the final expression of Eq.(26) and retaining the the zero order term and the term proportional to $\sqrt{c}$, only.

Some later, Redlich & Rosenfeld [50, 51] deduced an expression for the parameter K in Eq.(27) by combining the Nernst & Drude concept [48] with the Debye-Hückel theory of electrolyte solutions [79]. In Section 5, we will present the Redlich & Rosenfeld analytical expression supplemented by a discussion on it.

Owen & Brinkley [52, 54] criticized both the Mason limiting law given by Eq.(27) and the prediction of K made by Redlich & Rosenfeld [50, 51]. Instead, Owen & Brinkley suggested a more complex extrapolation formula which contained an additional parameter interpreted by the authors as the closest approach distance between the ions. Later, Redlich & Meyer [58] advocated Eq.(27) and their expression for K by demonstrating that the asymptotic behavior of the Owen & Brinkley function at $c \to 0$ coincides with Eq.(27) with K predicted in refs. [50, 51]. In ref. [58], Redlich & Meyer also proposed a correction of extrapolation function given by Eq.(27) by adding the third term on the right hand side of Eq.(27). The added term is assumed to be proportional to the electrolyte concentration, $c$, with a coefficient used as an adjustable parameter.

The density measurements data are processed with the help of Eq.(27), and the obtained dependencies, $v^*(c)$, are fitted by using the Mason- Redlich & Meyer function or/and the Owen & Brinkley one. Such a scheme has been conducted by different authors [49-78] for obtaining the adjustable parameters, in particular, the partial molar volumes of electrolytes at infinite dilutions, $v^*(0) = v(0)$. The values of the latter quantity are well documented for different electrolytes. For a given electrolyte, within the range of $c < 1\ mol/l$, the value of $v(c)$ reported in refs.[17-48] changes by about 10% compared to $v(0)$. Below, we briefly summarize the major data and regularities we found in the literature concerning $v(0)$ for different electrolytes.

For inorganic electrolytes, the reported partial molar volumes, $v(0)$, were found to be negative and positive, and vary within the range $-12$ to $62\ (cm^3/mol)$ [49, 54, 56, 60-65, 72-74, 76]. A trend is observed: for a given series of electrolytes having the same sets of ion charges, the partial molar volume increases with increasing the molar mass. However, in some cases, this trend is violated. Next, we survey some experimental data reported on different electrolyte types

*The 1:1 electrolytes.* With several exceptions ($NaOH$, $KOH$, $LiOH$, $KF$, $AgF$), the partial molar volumes have positive value that noticeably exceed the sum of crystallographic volumes of the respective ions. Except for the series including an acid and respective Lithium and Sodium salts, the value of $v(0)$ increases with increasing the electrolyte molar mass. The partial molar volumes lie between $-6.8\ cm^3/mol$ ($NaOH$) and $57.7\ cm^3/mol$ ($CsI$) [54, 56, 60, 62-64]

*The 2:1 electrolytes.* According to the available data (we did not find the information about the respective bases having low solubility), the partial molar volumes have positive values that increase with increasing the molar mass from $15.3\ cm^3/mol$ ($MgCl_2$) to $42.5\ cm^3/mol$ ($Pb(NO_3)_2$). Interestingly, very often, 2:1 electrolytes have noticeably smaller $v(0)$ than 1:1 electrolytes having lower molar mass. For example $Pb(NO_3)_2$ has an appreciably larger molar mass than $CsI$ and a lower partial molar volume. [54, 56, 63-65]

*The 1:2 electrolytes.* All the partial molar volumes found in the literature are positive. In the series consisting of Sulfuric Acid and sulfates of Alkali Metals, $H_2SO_4$ ($14.0\ cm^3/mol$); $Li_2SO_4$ ($12.2\ cm^3/mol$); $Na_2SO_4$ ($11.5\ cm^3/mol$); $K_2SO_4$ ($31.9\ cm^3/mol$) and $Cs_2SO_4$ ($56.7\ cm^3/mol$), with increase of the molar mass, the values of $v(0)$ either increase or remain nearly constant. [54, 56, 64]

<u>*The 2:2 electrolytes.*</u> In all the found examples, $MgSO_4$; $ZnSO_4$, $FeSO_4$, $CoSO_4$, $NiSO_4$, the partial molar volume turns out to be always negative. The volume absolute values are either smaller or slightly exceeding $10\ cm^3/mol$. [56, 64]

*The 3:1 electrolytes.* The partial molar volumes are positive and vary within a wide range which, for the found data, lies between $7.7\ cm^3/mol$ ($LuCl_3$) and $93.4\ cm^3/mol$ ($La(ClO_4)_3$). As it could be expected, $v(0)$ is mostly defined by the molecular mass of anion taking close values for salts having a common anion and different cations that belong to different parts of the periodic table. For example, $AlCl_3$ ($12.9\ cm^3/mol$) and $EuCl_3$ ($12.1\ cm^3/mol$); $Al(NO_3)_3$ ($43.0\ cm^3/mol$) and $Eu(NO_3)_3$ ($46.6\ cm^3/mol$) etc. [56, 61]

The above listed partial molar volumes of inorganic electrolytes demonstrate that the terms of order of $O(cv)$ might give noticeable contributions into the measured quantities for sufficiently concentrated solutions, For example, even $MgCl_2$ close to its solubility limit (about $6\ mol/l$) provides $cv \cong 0.09$. Another example $Al(NO_3)_3$ for which the estimations show $cv \cong$ 0.15 close of the solubility limit (about $3\ mol/l$). One can expect even higher values, $c\mathrm{v} \geq 0.2$, for the salts of Rear Earth Metals.

Definitely, the terms of order of $O(c\mathrm{v})$ are noticeable for some electrolytes containing organic ion since the partial molar volumes were found in the literature to be within the range 100 to 300 $(cm^3/mol)$ [66-70, 75, 77, 78]. Accordingly, the achievable volume fraction, $c\mathrm{v}$, sometimes can reach value of about 0.3. As well, such high volume fractions can be observed when a charge stabilized colloid is considered as electrolyte [80].

The role of such terms will be discussed in the end of Section 4

### 3.2 Discussion on thermodynamic fluxes in Kedem- Katchalsky-Michaeli theory

In the present section, we will focus on KKM theory [30,31] which was briefly discussed in Section 1. While considering membrane transport of binary electrolyte solution components, the authors of KKM theory transformed the EPF, $W$, by choosing such vectors of the Thermodynamic Forces and Fluxes, **X** and **Y**, that contain three components each and satisfy Eq.(2):

$$\mathbf{X} = \{\Delta\Pi,\ \Delta p - \Delta\Pi,\ \Delta E\} \quad (a)$$
$$\mathbf{Y} = \{J_s/c,\ J_v,\ I\} \quad (b) \tag{28}$$

where $I$ is the transmembrane electric current given by Eq.(7); $J_s$ and $J_v$ are referred to as the transmembrane electrolyte flux and volume flow in [30,31].

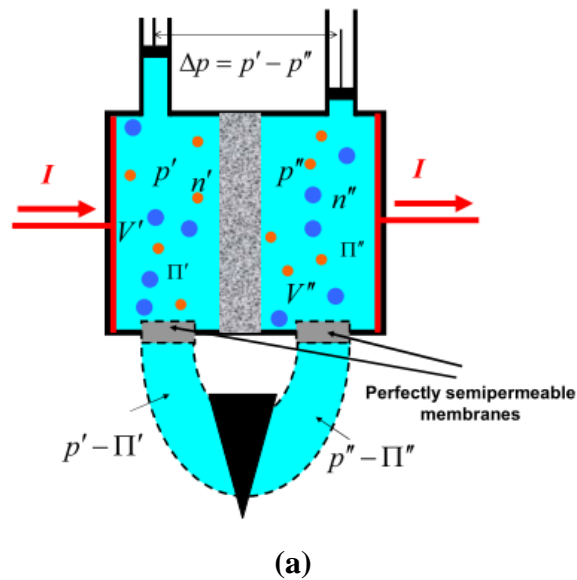

(a)

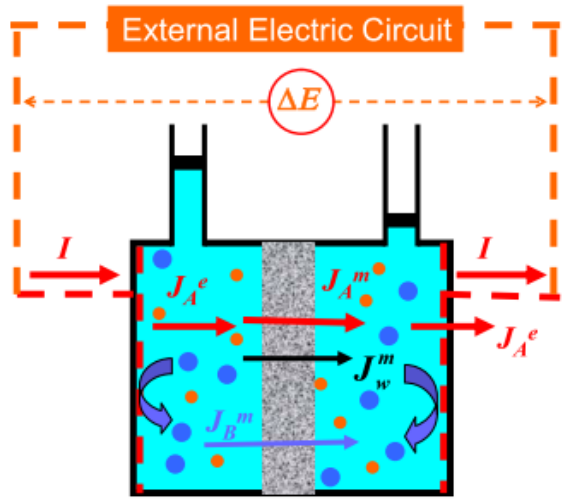

**(b)**

**Fig. 4 Illustrations to the discussion on physical meaning of some Thermodynamic Fluxes and Forces in the KKM theory: (a) Virtual Osmotic Pressure difference, ΔΠ; (b) the voltage, ΔE; the ion and solvent fluxes, $J_{A,B}^{m,e}$ and $J_w^m$ .**

Later in this Section, we will show that such an interpretation is associated with an approximation. In Section 5, we will estimate the range of parameters where such an approximation yields a good description.

Although the present communication deals with thermodynamic fluxes, next, we shortly comment the KKM set of thermodynamic forces given by Eq.(28a).

The quantity $\Delta p$ ($Pa$) in Eq.(28a) signifies the pressure difference across the membrane; $c\left(mol/m^3\right)$ is the solute concentration per unity of the solution volume in the solutions adjacent to the membrane. Note that the concentrations in the adjacent solutions may differ from each other. In refs.[30,31], the authors propose to use a mean solute concentration as the concentration $c$. In our opinion, using any concentration, which lies between those attributed to the adjacent solutions, does not lead to a higher error than the error originating from the use of linear approximation which is inherent in the LIT approach.

Two other quantities on the right hand side of Eqs.(28a), $\Delta\Pi$ and $\Delta E$, have non-trivial physical meanings In Eq.(28a), $\Delta\Pi$ is the transmembrane difference between the intensive thermodynamic parameter, $\Pi$, to which we will refer as the Virtual Osmotic Pressure (VOP) to distinguish it from the actual osmotic pressure. It should be stressed that $\Pi$ is not the pressure although it has the pressure dimension. However, for a given solution, the VOP can be determined by measuring the pressure difference in a special experiment where the solution is equilibrated with a pure solvent across a perfectly semipermeable membrane. The pressure difference across such an ideal membrane will yield the intensive parameter $\Pi$ attributed to a given solution.

The above meaning of VOP is illustrated in Fig.4a showing a U-pipe whose elbows contain the pure solvent and are separated by a plug. While bringing the elbows across two perfectly semipermeable membranes in contact with the solutions separated by the membrane under consideration, nothing changes in the cell shown in Fig.4a. Simultaneously, the pressure differences across the perfectly semipermeable membranes are established to be $\Pi'$ and $\Pi''$. Accordingly, the pressure difference across the plug in the U-pipe, $\Delta p_{plug}$, becomes equal to one of the thermodynamic forces in Eq.(28a), $\Delta p_{plug} = \Delta p - \Delta\Pi$.

Thus, when the plug is supplied with means of sensing the pressure differences, such a U-pipe could serve as a sensor for direct measuring the thermodynamic force $\Delta p - \Delta\Pi$ represented in the KKM equation set given by Eq.(28a). In particular, while maintaining zero value of the applied pressure difference, $\Delta p = 0$, such a device gives the first thermodynamic force in Eq.(28a), $\Delta\Pi$.

Another hypothetical set up suggested in [30,31] for interpreting the Thermodynamic Force, $\Delta E$, is sketched by the red dashed lines in Fig. 4b. Remarkably, it coincides with the VEC (B) shown in Fig 3 where the electrode transport numbers are given by the respective condition from Eq.(9). The quantity $\Delta E$ is named in [30,31] as the Electromotive Force and is interpreted there as the voltage which would exist between two external terminals of the VEC electrodes shown in (Fig.4b). It should be noted that the name Electromotive Force seems to be somewhat misleading terminology since, usually, it signifies the Open Circuit Voltage whereas, in Eq.(28a), one deals with a voltage in the presence of electric current through the cell.

Importantly, Eq.(9) implies two options of choosing VEC. The latter means that, strictly speaking, the KKM theory suggests to use not a unique vector of Thermodynamic Fluxes, but any of two vectors associated with blocking ions $A$ or $B$, $\mathbf{Y}^{(A)}$ or $\mathbf{Y}^{(B)}$ ($\mathbf{Y}^{(A,B)}$). Consequently, one can rewrite Eq.(28b), as

$$\mathbf{Y}^{(A,B)} = \left\{ J_s^{(A,B)}, J_v^{(A,B)}, I \right\} \tag{29}$$

where the electric current, $I$, given by Eq. (7) is a common component of both the versions of vector, $\mathbf{Y}^{(A,B)}$, whereas, in the general case, the first and second components differ: $J_s^{(A)} \neq J_s^{(B)}$ and $J_v^{(A)} \neq J_v^{(B)}$. While taking into account such a dualism inherent in the KKM theory, one can rewrite the expressions suggested in refs.[30,31] for the first and second components of the flux vector as

$$\begin{cases} J_s^{(A,B)} = J_{A,B}^m / \nu_{A,B} \\ J_v^{(A,B)} = J_s^{(A,B)}\mathrm{v} + J_w \mathrm{v}_w \end{cases} \tag{30}$$

Note that Eq.(30) comprises two equation sets. The first equation set is written with the superscripts (A) and the subscript $A$, to address the VEC blocking the ion $A$. The second equation set is attributed to the VEC blocking ion $B$, and thus one should use the superscript (B) and the subscript $B$.

A comparison of Eq.(30) with Eqs.(13) and (20) leads to the expected result

$$J_s^{(A,B)} = \left( J^* \right)_{\substack{t_{B,A}^e=1 \\ t_{A,B}^e=0}} \tag{31}$$

$$J_v^{(A,B)} = \left( J_v^* \right)_{\substack{t_{B,A}^e=1 \\ t_{A,B}^e=0}} \tag{32}$$

Thus, for any of two versions of the VEC, the KKM electrolyte and volume flows coincide with the apparent electrolyte and volume flows introduced in Sections 2 and 3.

Recall that the above apparent fluxes do not describe the purely membrane transfer, since they also take into account the electrode processes. The only exception is the regime of zero current. At $I=0$, the KKM vectors coincide, $\left(\mathbf{Y}^{(A)}\right)_{I=0} = \left(\mathbf{Y}^{(B)}\right)_{I=0} = \left(\mathbf{Y}\right)_{I=0}$. The latter becomes clear from the following sets of equalities

$$\left( J_s^{(A)} \right)_{I=0} = \left( J_s^{(B)} \right)_{I=0} = \left( J^m \right)_{I=0} \tag{33}$$

$$\left( J_v^{(A)} \right)_{I=0} = \left( J_s^{(A)}\mathrm{v} + J_w\mathrm{v}_w \right)_{I=0} = \left( J_s^{(B)}\mathrm{v} + J_w\mathrm{v}_w \right)_{I=0} = \left( J_v^{(B)} \right)_{I=0} = \left( J_v^m \right)_{I=0} \tag{34}$$

The chain of equalities given by Eq.(33) follows from Eqs.(14), whereas Eq.(34) is obtained while taking into account Eqs.(14) and (20). Thus, the first and second and components of the vector $(\mathbf{Y})_{I=0}$ yield the actual transmembrane electrolyte and volume flows, respectively.

When the electric current is not zero, the equality of the KKM vectors, $\mathbf{Y}^{(A)} = \mathbf{Y}^{(B)}$, is violated since $J_s^{(A)} \neq J_s^{(B)}$ and thus $J_v^{(A)} \neq J_v^{(B)}$.

To illustrate the regularities discussed above, we consider example of $MgCl_2$ solution whose components are transported through a membrane. The illustrations given in Figs.5a and 5b show two different versions of the VEC that are chosen for the same set of transmembrane fluxes as well as two different electrode couples. According to Eq.(13), when the electrode block transport of $Cl^-$-ions, the apparent electrolyte flux is given as $J^* = J_s^{(Cl^-)} = J_{Cl^-}^m / 2$. The electrode flux of $Mg^{2+}$-ion is determined from the continuity condition for the electric current, $I$, $J_{Mg^{2+}}^e = J_{Mg^{2+}}^m - J_{Cl^-}^m / 2$ (Fig.5a). When the electrodes block transport of $Mg^{2+}$-ions, the apparent electrolyte flux is given as $J^* = J_s^{(Mg^{2+})} = J_{Mg^{2+}}^m$. The electrode flux of $Cl^-$-ion is $J_{Cl^-}^e = -2J_{Mg^{2+}}^m + J_{Cl^-}^m$ (Fig.5b).

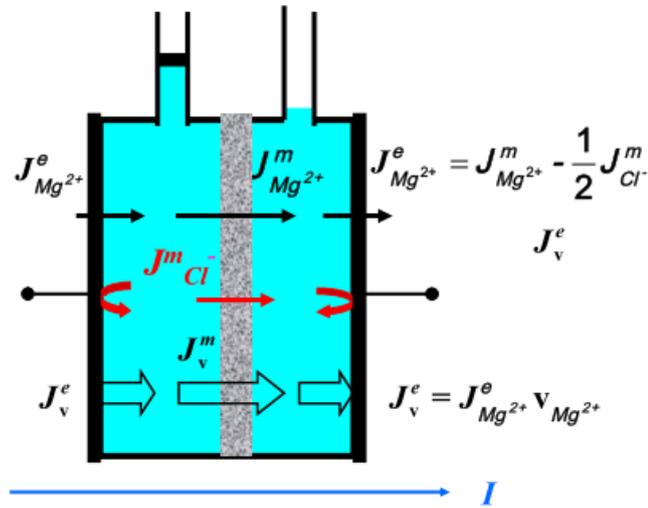

**(a)**

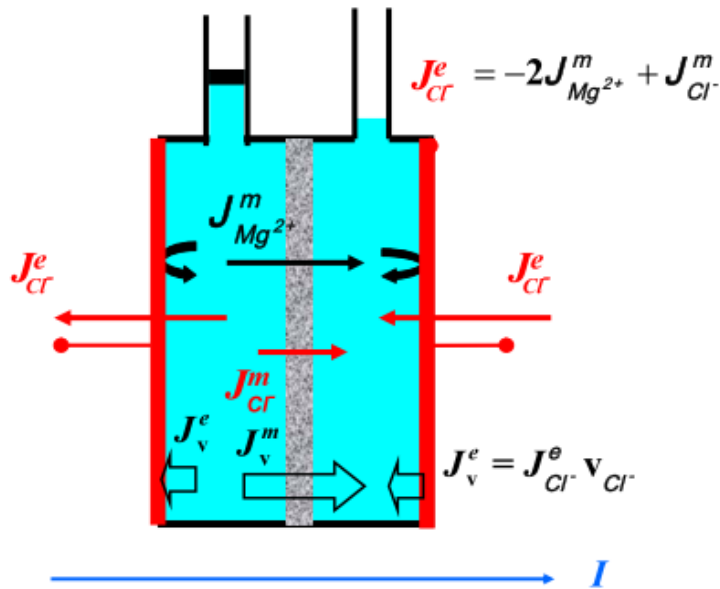

**(b)**

**Fig.5. Two versions of the Kedem Katchalsky-Michaeli couple of the Hypothetical Electrodes for the case of *MgCl₂* solution: the Hypothetical Electrodes block transport of *Cl⁻* - ions (a) or *Mg²⁺*- ions (b)**

The above example (Figs. 5 a and b) enables us to see that each of two KKM Thermodynamic Fluxes, $J_v^{(Cl^-)}$ and $J_v^{(Mg^{2+})}$, differs from the transmembrane volume flow by the quantity $J_v^e$ which gives a hypothetical out- or inward "leakage" of volume due to the electrochemical reactions involving, respectively, $Mg^{2+}$- or $Cl^-$-ions. Accordingly, for the VEC shown in Fig.5a, $J_v^e = J_{Mg^{2+}}^e \text{v}_{Mg^{2+}}$, and, for those shown in Fig.5b, $J_v^e = J_{Cl^-}^e \text{v}_{Cl^-}$. In two latter expressions, we introduce two quantities, $\text{v}_{Mg^{2+}}$ and $\text{v}_{Cl^-}$, that are referred to as the partial molar volumes of the ions. In the above relationships, these quantities serve as coefficients interrelating the rates of electrochemical withdrawal (delivering) $Mg^{2+}$ ($Cl^-$) -ions from (to) the solution and the rate of respective solution volume changes.

In the KKM theory [30,31], there are two other approximations that have been made while obtaining $\Delta\Pi$ and $\Delta E$. The first of them amounts to omitting the terms of order $O(c\text{v})$ in the expression for $\Delta\Pi$. As shown in the end of section 3.1, such terms might be noticeable for sufficiently concentrated solution. The second of abovementioned approximations is that the terms of order of $O(\text{v}_{A(B)}\Delta p)$ have been omitted in the expression for $\Delta E$. The analysis conducted in the present Section shows that the identification of $J_v$, which is given by Eqs.(30), as the transmembrane volume flow leads to certain error of order of the abovementioned term, $J_{A(B)}^e \text{v}_{A(B)}$.

For the ranges of parameters where the approximations work well, the KKM theory gives perfect and elegant description of membrane transport. The elegance is that the KKM theory deals with a purely thermodynamic analysis and does not employ extra-thermodynamic assumptions. However, to determine the frameworks for the KKM theory applicability and to extend it, one should know the meaning and properties of the partial molar volumes of different ions, $\text{v}_{A(B)}$.

In Section 4, we will present a detailed discussion on that. Now, we only mention that, in contrast with the akin quantity attributed to electrolytes, v, and discussed in Section 3.1, $\text{v}_{A(B)}$ is not measured directly and is extracted from certain experimental data by using extra-thermodynamic assumptions. Consequently, for $I \neq 0$, refusing from the KKM approximation unavoidably requires using extra-thermodynamic assumptions regarding the quantity $\text{v}_{A(B)}$ ($\text{v}_{Mg^{2+}}$ and $\text{v}_{Cl^-}$, in the example illustrated by Fig. 5).

**4. Partial molar volumes of ions and transmembrane volume flow**

A difficulty in describing the transmembrane flux in the presence of transmembrane electric current, $I \neq 0$, originates from the fact that the electrolyte ions are transported through membrane in non-stoichiometric amounts. Therefore, it is required to introduce a phenomenological quantity describing the changes of adjacent solution volume per unity of added amount of an individual ion, not the electrolyte. In the respective definition, the amount of the other ion should remain unaltered. Formally, the respective thermodynamic quantity, which is referred to as the partial molar volume of ion, can be introduced with the help of the thought experiment discussed next.

*4.1 Schemes of obtaining the partial molar volume of an ion*

Similarly to the scheme employed in Section 3.1, Eqs.(18)-(22), for defining the partial molar volume of electrolyte, we consider the system equation of state in the form representing the system volume as a function of several parameters. However, in contrast with Eq.(16), the present equation of state does not employ the electroneutrality condition given by Eq.(6a) and thus allows ion amounts, $n_A$ and $n_B$, to change independently from each other. As the proposed consideration admits appearance of electric charge in the system, it is necessary to introduce electric potential, $\Phi$, as one of the parameters that define the system volume. Hence, the analogy of Eq.(16) can be written as

$$V = V(T, p, \Phi, n_A, n_B, n_w) \tag{35}$$

The electric potential, $\Phi$, has complex meaning in Statistical Physics and, in a thought experiment under consideration, can be regulated/measured with the help of a special electrode.

Now, we define the partial molar volume of ions as

$$\mathrm{v}_A = \left(\frac{\partial V}{\partial n_A}\right)_{P,T,n_B,n_W,\Phi} \quad (a) \qquad \mathrm{v}_B = \left(\frac{\partial V}{\partial n_B}\right)_{P,T,n_A,n_W,\Phi} \quad (b) \tag{36}$$

Thus, in the discussed thought experiment, the partial molar volume of ion $A$ (or $B$), $\mathrm{v}_{A(B)}$ is determined by considering the small change of solution volume, $dV$, due to an addition of small portion of the ion $A$ (or $B$), $dn_{A(B)}$, while maintaining constant pressure, $P$, temperature, $T$, amount of other ion $n_{B(A)}$ and electric potential, $\Phi$. The latter quantity can be held constant by using external electric charges that induce in the solution the electric potential opposite to that produced by the added charge, $Fz_{A(B)}dn_{A(B)}$.

In the literature, we did not find any attempt of practical implementing the hypothetic thermodynamic scheme outlined above. Instead, in a great number of publications the partial molar volumes of individual ions are determined by using the measured partial molar volumes for a set of electrolytes with a common ion and applying the additivity rule [55]

$$\mathrm{v} = \nu_A \mathrm{v}_A + \nu_B \mathrm{v}_B, \tag{37}$$

When the partial molar volume of the common ion is known, using Eq.(37) enables one to determine the volume for all other ions represented in the abovementioned set. Usually, by assuming certain value for partial molar volume of $H^+$-ions, $v_{H^+}$, the volumes are reconstructed for all other ions. It is often assumed that $v^{H^+} = 0$. Another value, $v^{H^+} \cong 5\, cm^3/mol$, was proposed by Millero [63] on the basis of extra-thermodynamic reasons.

The above scheme based on Eq.(37), does not contradict the thermodynamic definition based on Eqs.(36). At the same time, one cannot exclude the situations when the additivity rule given by Eq.(37) is violated. Hence, we will consider Eq.(37) as an assumption whose validity has been confirmed for many electrolytes.

*4.2 Transmembrane volume flow*

While inspecting Fig.3 and using the definition given by Eq.(37), one can represent the volume flow transferred through the membrane, as

$$J_v^m = J_A^m v_A + J_B^m v_B + J_w^m v_w \qquad (38)$$

Now, by combining Eqs.(5), (7), (37) and (38), we arrive at an expression which is symmetric with respect to both the ions

$$J_v^m = Jv + J_w^m v_w + \frac{I}{F}\frac{v_A - v_B}{z_A - z_B} \qquad (39)$$

where $J$ is represented, as

$$J = \frac{J_A^m + J_B^m}{v_A + v_B} \qquad (40)$$

The introduced quantity, $J$, can be interpreted in terms of the KKM theory. By combining Eqs.(30) and (40), one obtains

$$J = \frac{J_s^{(A)} v_A + J_s^{(B)} v_B}{v_A + v_B} \qquad (41)$$

Thus, the quantity $J$ is a sort of average of two versions of the KKM electrolyte flux. For $I = 0$, according to Eqs.(33) and (41), $J$ coincides with the common KKM electrolyte flux, $\left(J^m\right)_{I=0}$.

Another interesting meaning of the quantity $J$ becomes clear while assuming that the VEC electrodes have transport numbers coinciding with the relative ionic strength of the respective ion in the solution, i.e., when $t_{A,B}^e = z_{A,B}^2 c_{A,B}/\left(z_A^2 c_A + z_B^2 c_B\right)$. By substituting the latter expression into the final expression of Eq.(12), after some transformations using Eqs.(5) and (6), we arrive at the right hand side of Eq.(40). Recall that Eq.(12) gives the apparent electrolyte flux, $J^*$, for an arbitrary set of electrode transport numbers, $t_{A,B}^e$. Thus, for the VEC with the electrodes having the abovementioned transport numbers, the quantity $J$ coincides with the apparent electrolyte flux, $J^*$. One can suggest several purely theoretical constructs that could be considered as models for such electrodes. However, we did not find practical implementations of them.

The transmembrane and KKM volume flows can be interrelated by combining Eqs.(30), (38) and (39)

$$J_v^m = J_v^{(A)} + \frac{I}{Fz_B}v_B = J_v^{(B)} + \frac{I}{Fz_A}v_A \qquad (42)$$

Alternatively, Eq.(42) could easily be derived by using VEC blocking either $A$ or $B$ ions. Such a derivation amounts to adding the terms describing the volume changes due to the electrochemical reactions, $J_v^e = Iv_{B,A}/Fz_{B,A}$, to the KKM volume flows, $J_v^{(A,B)}$. The added terms describe contribution of the ion which is not blocked. Note, however, the derivation represented above Eq.(42) was dealing with the membrane, only, an did not use VEC concept. It is a clear illustration of the fact that using the VEC concept is helpful in derivations, but is not a necessary step in the analysis.

Let us now come back to the expressions that are given by Eqs. (39) and (40) for describing the transmembrane volume flow, $J_v^m$. When $I \neq 0$, the right hand side of Eq.(39) contains the term proportional to the difference between the individual ion partial molar volumes. To judge about the importance of the latter term on the right hand side of Eq.(39), next, we present a brief survey of literature data on the partial molar volumes of different ions and a discussion on the physical effects defining the observed values.

*4.3 Partial molar volumes of different ions. Electrostriction*

Three trends are observed for the partial molar volumes of inorganic ions at the infinity dilution limit, $v_{A,B}(0)$, [52-65, 71-74, 76, 77]:

a) for a series of chemically akin elements, with rising the molar mass, the respective ion volumes increase or remain approximately unaltered;

b) anions have positive partial molar volumes that increase with molar mass and are always larger than that of cations with the same absolute values of charge;

c) the volumes noticeably decrease with increasing the ion charge that leads to negative partial molar volumes of multi-charged cations.

Now, we will illustrate these trends by some literature data reported for different ions.

*Alkali Metals* [71]. With increasing the molar mass, the partial molar volumes increase from slightly negative to positive values. Such a regularity is observed in the series $Na^+$-, $K^+$-, $Rb^+$- and $Cs^+$-ions where $v_A(0)$ changes from $-6.62$ to $15.93$ $cm^3/mol$. As for the $Li^+$-ion, it somewhat violates this regularity since its negative partial molar volume $\left(-6.29\ cm^3/mol\right)$ is reported to be slightly more than that of $Na^+$-ion $\left(-6.62\ cm^3/mol\right)$.

*Alkaline Earth Metals* [71]. For the series $Mg^{2+}$-, $Ca^{2+}$-, $Sr^{2+}$- and $Ba^{2+}$-ions, with increasing the molar mas, the volumes increase, (i.e., decrease by the absolute values) from $-31.99$ to $-23.29 cm^3/mol$. The described trend is violated in the case of transition from $Ca^{2+}$- to $Sr^{2+}$-ion for which there is a slight decrease of $v_A(0)$ from $-28.67$ to $-29.18 cm^3/mol$.

*Transition Metals* [54, 56, 63]. In the series $Mn^{2+}$-, $Fe^{2+}$-, $Co^{2+}$-, $Ni^{2+}$-and $Zn^{2+}$-ions, one observes negative values of a $v_A(0)$ and slight monotonous increase from $Fe^{2+}$ $\left(-25.3\ cm^3/mol\right)$ to $Zn^{2+}$ $\left(-22.1\ cm^3/mol\right)$. However, the $Mn^{2+}$-ion violates the trend since it has the lowest molar mass in the series and the smallest absolute value of the negative partial molar volume $\left(v^{Mn}(0)=-18.3\ cm^3/mol\right)$

*Rare Earth Metals* ($z_A=+3$) [61]. While considering the elements of the Lanthanide Series (numbers 57-71 in the Periodic Table), one observes a negative partial molar volumes. In this series, with increasing the molar mass, the volume slightly and non-monotonously varies between $-42.0\ cm^3/mol$, for $La^{3+}$-ion, to $-48.8\ cm^3/mol$, for $Lu^{3+}$-ion

*Halogens* ($z_A=-1$) [52, 56, 71]. In the series $F^-$, $Cl^-$, $Br^-$ and $I^-$-ions, all the partial molar volumes are positive and monotonously increase from $4.25$ to $41.63\ cm^3/mol$

*Polyatomic anions* [52, 56, 71] The series $NO_3^-$ $\left(29.3\ cm^3/mol\right)$; $ClO_4^-$ $\left(49.53\ cm^3/mol\right)$ and $SO_4^{2-}$ $24.8\ cm^3/mol$, illustrates all the trends listed under item (b).

Above, we summarized the existing experimental data on the infinity dilution limit which is approached by the partial molar volumes of different electrolytes and ions in aqueous solution. In the literature, the observed properties are explained through different mechanisms of the solvent reorganization by ions. In particular, the negative values of the ion partial molar volumes are explained trough the electrostriction produced by ions. The respective quantitative theory was proposed by Drude & Nernst who addressed the electrostriction in solvent due to the Columbic fields created by individual ions [48].

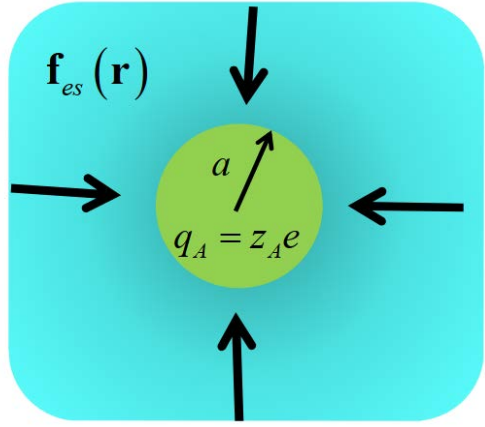

**Fig.6 Illustration for electrostriction effect produced by one ion**

In the Drude & Nernst theory [48], the solvent is considered to be a homogenous compressible perfect dielectric medium which surrounds an ion (Fig.6). The ion is modeled as a spherical charged particle having the radius, $a$, and the charge, $q_A = ez_A = Fz_A/N$, where $N \approx 6.02 \cdot 10^{23} \, mol^{-1}$ is the Avogadro number. The solvent is compressed due to the pressure gradient which is produced to compensate for the electrostriction force, $\mathbf{f}_{es}(\mathbf{r})$, [81] given as

$$\mathbf{f}_{es}(\mathbf{r}) = \frac{1}{2} d_w \left( \frac{\partial \varepsilon_w}{\partial d_w} \right)_T \nabla(E^2) \tag{43}$$

where $E = q_A / 4\pi\varepsilon_w r^2$ is the local electric field strength created by the ion; $\varepsilon_w$ is the solvent dielectric permittivity.

Such a compression results in a decrease of solvent volume which turns out to be proportional to the added amount of the ion $A$. For convenience of discussion, we represent the Drude & Nernst [48] final expression for the partial molar volume of an ion $v_A(0)$, which is rederived in Appendix 2, in terms of two parameters, $v_{D-N}(m^3/mol)$ and $v_{int}(m^3/mol)$, as

$$v_A(0) = v_{int}\left[1 - z_A^2 \left(\frac{v_{D-N}}{v_{int}}\right)^{4/3}\right] \tag{44}$$

Hereafter, we will refer to $v_{int}(m^3/mol)$ and $v_{D-N}(m^3/mol)$ as the intrinsic and Drude-Nernst molar volumes, respectively. In terms employed in the Drude-Nernst theory, the abovementioned parameters are expressed as

$$v_{int} = \frac{4\pi}{3} a^3 N \quad \text{(a)}$$

$$v_{D-N} = \frac{1}{6\sqrt{\pi}N} \left\{ \frac{3F^2}{2\varepsilon_w} \left[ \frac{\partial \ln(\varepsilon_w)}{\partial p} \right]_T \right\}^{3/4} \quad \text{(b)} \tag{45}$$

where $p$ is the pressure. Note that Eq.(45b) is written by using the SI convention.

While interpreting the length scale parameter $a$ as the ion crystallographic radius, we estimate it, roughly, as $a \approx 10^{-10} m$. Such an estimation is made with understanding that the reported crystallographic radii of inorganic ions may deviate from the above estimation not more than by a factor of about two [56]. Accordingly, by applying Eq.(45a) for such an ion, one obtains for its intrinsic molar volume $v_{int} \approx 2.52\ cm^3/mol$.

The Drude-Nernst molar volume is completely defined by the dielectric properties of solvent. To evaluate $v_{D-N}$, we substitute in Eq.(45b) $\varepsilon_w = 6.9 \cdot 10^{-10} F/m$; $F = 9.65 \cdot 10^4 C/mol$ and the dielectric compressibility value taken from refs. [50,51,58], $\left[\partial \ln(\varepsilon_w)/\partial p\right]_T = 4.71 \cdot 10^{-10} Pa^{-1}$. Finally, we obtain the Drude-Nernst molar volume, $v_{D-N} \approx 3.64\ cm^3/mol$.

The partial molar, $v_A(0)$, given by Eq.(44), can be represented as a sum of two terms. The first one is given by the intrinsic partial volume, $v_{int}$, which is always positive. Expectedly, the second term is always negative because it describes the electrostriction which results in the compression of solvent. Consequently, the sign of partial molar volume becomes negative when the negative part prevail by absolute value, i.e., for sufficiently small ions having sufficiently high charges.

For three absolute values of ion charge, the curves plotted in Fig.7 display the behavior discussed above and described by Eq.(44).

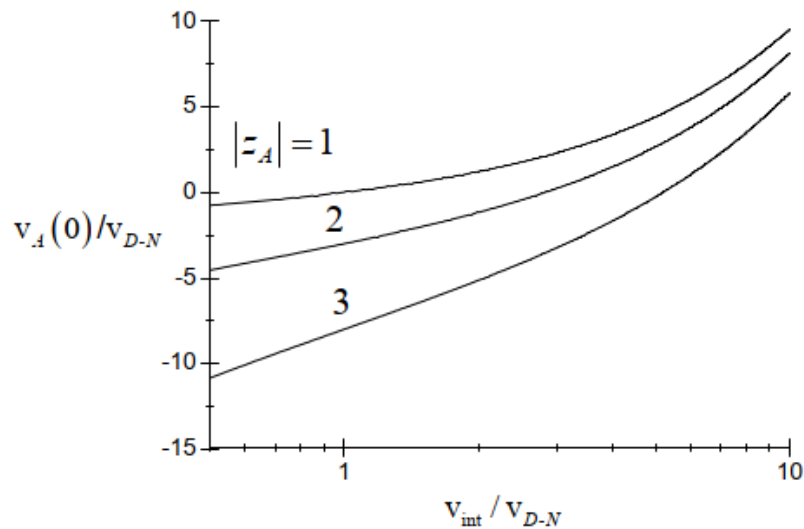

**Fig.7 Normalized partial molar volume of an ion as a function of the normalized intrinsic volume, Eq.(44)**

Now, we inspect the graphs and the experimental data on $v_A(0)$ listed above. While taking into account the estimations, $v_{D-N} \approx 3.64 \; cm^3/mol$, we arrive at the following conclusions. For cations, one can roughly fit the experimental data by choosing such values of $v_{int}$ in Eq.(44) that correspond to the values of $a$ that are slightly less than $10^{-10} m$. For anions, one should assume appreciably larger $a$. The latter seems to correspond to the first Pauling rule [82].

To address such behavior of anions in terms of the Drude & Nernst general approach, Mukerjee [57] and Glueckauf [59] considered voids that exist around the ions due to the discontinuous molecular structure of solvent. Accounting for that was conducted by introducing a void shell which envelopes the "bare" ion and thus separates it from the solvent. Accordingly, the intrinsic molar volume is interpreted as the sum of total volumes inside external shell boundaries for 1 *mol* of ions. Clearly, this volume is more than that of "bare" ions. Remarkably, the analysis of the Mukerjee & Glueckauf model [57, 59] leads to the result given by Eqs.(44) and (45b) but with the parameter $v_{int}$ defined by the volumes inside the void shell boundaries.

There are different more sophisticated models in the literature. Millero [62, 63] took into account the existence of disordered water molecules around the ion. Marcus [72, 73] considered the dielectric non-linearity in the vicinity of ion where the electric field is extremely strong. One should also mention the study of Couture and Laidler [56] where the authors, by analyzing the experimental data of Owen and Brinkley [52, 54], suggested interpolation formulas that represent the partial molar volumes of ions as functions of their crystallographic radii and charges.

As we already mentioned in Section 3.2, the electrostriction mechanism was also analyzed by Redlich & Rosenfeld [50,51,58] for addressing the empirically established high dilution behavior of the apparent electrolyte volume as a function of concentration given by Eq.(27), $v^*(c)$. In this theory, instead of the Columbic field around an ion used by Drude & Nernst, the authors considered the Debye field thereby taking into account the screening effect [79]. The Debye screening effect results in two trends. On the one hand, the screening charge is attracted to a given ion to produce a contribution into the excess pressure around the ion that additionally compresses the solvent. On the other hand, the screening makes the electric field weaker than the purely Columbic field thereby decreasing electrostriction compared to the Drude & Nernst prediction.

In the terms employed above, the constant $K \left( m^{9/2}/mol^{3/2} \right)$ obtained by Redlich & Rosenfeld [50,51,58] for the limiting "square root" law given Eqs.(27) can be written as

$$K = \left(\frac{\nu_A z_A^2 + \nu_B z_B^2}{4\varepsilon_w}\right)^{3/2} \frac{F^3}{N\pi\sqrt{RT}} \left[\frac{1}{\varepsilon_w}\left(\frac{\partial \varepsilon_w}{\partial p}\right)_T - \frac{1}{3d_w}\left(\frac{\partial d_w}{\partial p}\right)_T\right] \qquad (46)$$

In Appendix A2.3, we rederived Eq.(46). Note also that, in contrast with the original result [50,51,58], Eq.(46) is represented in the terms of the SI convention.

### *4.4 Estimations of ion transfer contribution into the volume flow*

Let us estimate whether the terms, $I v_{B,A} / F z_{B,A}$, correcting the KKM volume flow, $J_v^{(A,B)}$, are important for obtaining the actual transmembrane flow, $J_v^m$, in Eq.(42). First, we compare the abovementioned correcting terms with the KKM terms expressing contribution of ions $J_s^{(A,B)} v = J_{A,B}^m v / \nu_{A,B}$ by introducing a criterion, $\alpha$, which yields the ratio of the correcting to KKM term. While choosing for certainty the VEC, which blocs the ion $A$, one obtains the following chain of equalities in

$$\alpha = \left|\frac{I v_B}{F z_B J_s^{(A)} v}\right| = \left|\frac{I v_B \nu_A}{F z_B J_A^m v}\right| = \left|\frac{I v_B \nu_B}{F z_A J_A^m v}\right| \qquad (47)$$

For obtaining the final expression in Eq.(47), we used Eq.(5).

Consider now the ratio $|I|/F|z_A J_A^m|$ in the latter expression of Eq.(47). For $I \to 0$, the ratio approaches zero, always, except for the cases when the transmembrane flux of the blocked ion, $A$, is proportional to the electric current, $J_A^m \cong I$. In the latter case, the KKM volume flow also approaches zero when $I \to 0$ such that the abovementioned ratio, $|I|/F|z_A J_A^m|$, approaches a finite value. The proportionality, $J_A^m \cong I$, always occurs when the directions of fluxes are defined by electro-migration. In such cases, the transmembrane fluxes $J_A^m$ and $J_B^m$ are oppositely directed and, thus, $|I| > F|z_A J_A^m|$, as it is clear from Eq. (7).

The multiplier $|v_B \nu_B / v|$ represented in the final expression of Eq.(47) can be greater than unity for inorganic electrolytes composed by multi-charged cations having negative partial molar volume, as stated in Sections 4.3. For example, while using the data on $MgCl_2$ listed in Sections 3.1 and 4.3, one obtains that $|v_{Mg^{2+}} / v| \approx 2$. Thus, at least, for all the above discussed cases, $\alpha > 1$, that defines importance of the term omitted in the KKM when ions bring noticeable contribution into the volume flow.

Let us now consider Eq.(39) to evaluate whether the latter term on the right hand side yields a measurable contribution into the transmembrane volume flow. Note that the term in question depends on the partial molar volumes of ions, $v_{A,B}$, and disappears when they are equal. We will evaluate the abovementioned term by introducing the change of apparent zeta potential, $\Delta\zeta^*$, due to the presence of respective contribution into the transmembrane volume flow. The apparent zeta potential is obtained while normalizing the transmembrane volume flow by the Smoluchowski factor $\chi = \varepsilon_w I / \eta g$ where $\eta$ is the solvent viscosity, and $g$ is the conductivity of the equilibrium solution [16,17]

$$\left|\Delta\zeta^*\right| = \left|\frac{I}{F\chi}\frac{v_A - v_B}{z_A - z_B}\right| = \left|\frac{\eta g}{F\varepsilon_w}\frac{v_A - v_B}{z_A - z_B}\right| > \frac{\eta}{\varepsilon_w}\frac{|v_A - v_B|}{v_A + v_B}\frac{Fc|z_A z_B|(|z_A| + |z_B|)D_{\min}}{RT} \qquad (48)$$

For obtaining the final inequality of Eq.(48), the conductivity was estimated as $g < F^2 c |z_A z_B|(|z_A| + |z_B|) D_{\min} / RT$, where $D_{\min}$ is the diffusion coefficient of the slowest ion.

By introducing the hydrodynamic radius of the slowest ion with the help of the Stokes-Einstein formula, $a^h_{slowest} = RT / 6\pi\eta D_{\min} N$, and using Eq.(5), one obtains from the latter inequality of eq.(48)

$$\left|\Delta\tilde{\zeta}^*\right| = \frac{\Delta\zeta^* F}{RT} > \frac{2\beta|z_A z_B|}{3a^h_{slowest}} c |v_A - v_B| \qquad (49)$$

where $\tilde{\zeta} = \zeta F / RT$ and the value of the Bjerrum length is $\beta = F^2 / 4\pi\varepsilon_w RT N \approx 7 \cdot 10^{-10} m$.

For diversity, we will evaluate $\left|\Delta\tilde{\zeta}^*\right|$ attributed to $FeCl_3$ and take $a^h_{slow}$ from the diffusion coefficient given in handbook [75] to see that $2\beta / 3a^h_{slowest} \approx 1$. While considering an aqueous solution $FeCl_3$ ($v_{Cl^-} \approx 24\ cm^3/mol$ and $v_{Fe^{3+}} \approx -44\ cm^3/mol$, [59]), we obtain the following rough estimation $\left|\Delta\tilde{\zeta}^*\right| \geq 0.2\ c(mol/l)$ For concentrations of about $5\ mol/l$, which is close to solubility limit, $\left|\Delta\tilde{\zeta}^*\right| \geq 1$.

When the electric current is passed through the membrane at zero concentration difference, one can see that the third term of the right hand side of Eq.(39) can be greater by magnitude than the first one. Such a situation definitely takes place when the ion fluxes are of migration origin and thus are directed oppositely. Both these terms describe the contribution of ions to the volume flow. Hence, the contribution of ions into the volume flow during electroosmosis can change the normalized value of the measured zeta potential within the estimated range 0.2-1. Note that it is of order or even more than the zeta-potentials attributed to the solvent transfer at such high concentrations.

## 5. Coupling between the transmembrane fluxes of different nature

Now, we will focus on a wide class of membrane transport phenomena listed in Section 1.3 for which imposing a transmembrane flux of one physical nature results in generating transmembrane fluxes of other natures. Such a coupling between the fluxes is defined by membrane properties that are described by a set of phenomenological coefficients. In the present section, we will give the definitions of these coefficients on the basis of different thought experiments. The definitions to be proposed will be based on conservation laws and linearity reasons and thus will not use the Onsager theorem.

For introducing the abovementioned kinematic coefficients we subdivide all the membrane cross-effects in three groups that, as discussed in Section 1.1, are associated with mechano-chemical, electrochemical and electromechanical (electrokinetic) transformation of free energy. Next, we introduce an important quantity which will be employed while defining the coefficients that describes the abovementioned coupling between the fluxes

### *5.1 Chemical fluxes*

As stated in Section 1, the basic property of membranes is the ability to provide transport of solution components in a proportion which differs from that taking place in the adjacent solutions. Accordingly, the major indicator of membrane separation effect is the changes of component concentrations in the adjacent solutions due to the transfer of them through the membrane.

To describe the total rate of ion concentration changes within any of compartments of the VEC shown in Fig.3, we introduce an important quantity to which we will refer as the chemical flux of electrolyte, $J^{ch}$. It is convenient to use different signs while defining this quantity for different compartments shown in Fig.3

$$\begin{aligned} \left(J^{ch}\right)' &= -V'\frac{dc'}{d\tau} \quad &\text{(a)} \\ \left(J^{ch}\right)'' &= V''\frac{dc''}{d\tau} \quad &\text{(b)} \end{aligned} \quad (50)$$

In a similar manner, we define the ion chemical fluxes, $J^{ch}_{A,B}$

$$\begin{aligned} V'\frac{dc'_{A,B}}{d\tau} &= -\left(J^{ch}_{A,B}\right)' = -\nu_{A,B}\left(J^{ch}\right)' \quad &\text{(a)} \\ V''\frac{dc''_{A,B}}{d\tau} &= \left(J^{ch}_{A,B}\right)' = \nu_{A,B}\left(J^{ch}\right)'' \quad &\text{(b)} \end{aligned} \quad (51)$$

Below, we will demonstrate, that, in the general case, the chemical fluxes attributed to different compartments differ from each other, $\left(J^{ch}\right)' \neq \left(J^{ch}\right)''$ and $\left(J^{ch}_{A,B}\right)' \neq \left(J^{ch}_{A,B}\right)''$.

The chemical fluxes introduced above can be expressed in the terms of apparent fluxes, $J^*$, $J^*_{A,B}$ and $J^*_v$, defined by Eqs.(11), (15) and (18). By combining these definitions with Eqs.(50) and (51), one obtains

$$\begin{aligned}\left(J^{ch}\right)' &= J^* - c'J^*_v \quad &\text{(a)} \\ \left(J^{ch}\right)'' &= J^* - c''J^*_v \quad &\text{(b)}\end{aligned} \qquad (52)$$

and

$$\begin{aligned}\left(J^{ch}_{A,B}\right)' &= J^*_{A,B} - c'_{A,B}J^*_v \quad &\text{(a)} \\ \left(J^{ch}_{A,B}\right)'' &= J^*_{A,B} - c''_{A,B}J^*_v \quad &\text{(b)}\end{aligned} \qquad (53)$$

In Eqs.(52) and (53), it is taken into account that, according to the definitions given by Eqs.(11), (15) and (18), the apparent fluxes, $J^*$, $J^*_v$ and $J^*_{A,B}$, attributed to different compartments of the cell shown in Fig.3 are equal. Thus, the difference between chemical fluxes originates from the differences in electrolyte or ion concentrations, i.e., $\left(J^{ch}\right)' \neq \left(J^{ch}\right)''$ and $\left(J^{ch}_{A,B}\right)' \neq \left(J^{ch}_{A,B}\right)''$ when $\Delta c = c' - c'' \neq 0$.

At the same time, very often, within the frameworks of LIT, the non-linear terms, including bilinear ones, are ignored. As it follows from Eqs.(52) and (53)

$$\begin{aligned}\left(J^{ch}\right)' - \left(J^{ch}\right)'' &= O\left(J^*_v \Delta c\right) \\ \left(J^{ch}_{A,B}\right)' - \left(J^{ch}_{A,B}\right)'' &= O\left(J^*_v \Delta c_{A,B}\right)\end{aligned} \qquad (54)$$

Thus, while using the LIT, we will consider a single chemical flux for each of the ions, $J^{ch}_A$ and $J^{ch}_B$, as well as for electrolyte, $J^{ch}$.

However, it should be stressed that the above identification of chemical fluxes is valid for sufficiently small $\Delta c$ and $J^*_v$. Accordingly, while analyzing non-linear regimes with the help of continuous version of the LIT of [37-46], the chemical flux turns out to be discontinuous.

By using Eqs.(10a) and (15), each of the ion chemical fluxes can be broken into the membrane and electrode parts

$$J^{ch}_{A,B} = \left(J^{ch}_{A,B}\right)^m - \left(J^{ch}_{A,B}\right)^e \qquad (55)$$

where

$$\begin{aligned}\left(J^{ch}_{A,B}\right)^m &= J^m_{A,B} - c_{A,B}J^m_v = J^m_{A,B} - \nu_{A,B}cJ^m_v \quad &\text{(a)} \\ \left(J^{ch}_{A,B}\right)^e &= \frac{It^e_{A,B}}{Fz_{A,B}} - c_{A,B}J^e_v = \frac{It^e_{A,B}}{Fz_{A,B}} - \nu_{A,B}cJ^e_v \quad &\text{(b)}\end{aligned} \qquad (56)$$

For the case of zero current, when, according to Eq.(14), the transmembrane and apparent fluxes coincide, $J^m = J^*$ and $J_v^m = J_v^*$, the VEC electrodes do not contribute to the mass exchange. Consequently, one can rewrite Eq.(53) and (54) in the terms of transmembrane fluxes as

$$\left(J^{ch}\right)'_{I=0} = \left(J^m\right)_{I=0} - c'\left(J_v^m\right)_{I=0} = \frac{\left(J_A^{ch}\right)'_{I=0}}{\nu_A} = \frac{\left(J_B^{ch}\right)'_{I=0}}{\nu_B} \quad \text{(a)}$$

$$\left(J^{ch}\right)''_{I=0} = \left(J^m\right)_{I=0} - c''\left(J_v^m\right)_{I=0} = \frac{\left(J_A^{ch}\right)''_{I=0}}{\nu_A} = \frac{\left(J_B^{ch}\right)''_{I=0}}{\nu_B} \quad \text{(b)}$$

(57)

The chemical fluxes introduced in the present section bear information about the rates of concentration changes that can directly be measured in the adjacent solutions. Therefore, the chemical fluxes are convenient quantities for using in studies of mechano- and electrochemical coupling between the fluxes. In the next two sections these types of couplings are considered in details.

*5.2 Mechano-chemical coupling*

In this Section, we will consider coupling between the volume flow and the chemical flux at zero current regime, $I = 0$. Under this condition, we consider equal electrolyte concentrations in the compartments shown in Fig.3, $c = c' = c''$. While imposing volume transfer through a membrane, one observes changes in the electrolyte concentrations in the adjacent solutions. According to Eqs.(50), the rates of such changes are defined by the chemical flux which takes a common value for both the compartments in the case under consideration. This common value is given by Eq.(57). Consequently, when the transmembrane flux coincides with the convective fluxes entering and leaving the membrane, $J^m = J_v^m c$, no concentration changes are observed since $\left(J^{ch}\right)^m = 0$. When $\left(J^{ch}\right)^m \neq 0$, one can observe concentration changes in the compartments.

It is convenient to introduce the electrolyte reflection coefficient, $\sigma$, defined as

$$\sigma = -\left(\frac{J^{ch}}{J_v^m c}\right)_{\substack{\Delta c=0 \\ I=0}} = \left(1 - \frac{J^m}{J_v^m c}\right)_{\substack{\Delta c=0 \\ I=0}} = \left(1 - \frac{J_{A,B}^m}{J_v^m c_{A,B}}\right)_{\substack{\Delta c=0 \\ I=0}} \quad (58)$$

Another coefficient, which is used for describing the mechano-chemical coupling between the fluxes, is referred to as the rejection, $\gamma$. The rejection is introduced for addressing the membrane separation effect for the schemes employed in baromembrane separation technologies when $\left(J^{ch}\right)' \neq \left(J^{ch}\right)''$. The rejection describes the rate of concentration changes in the compartment with the outward transmembrane volume flow (feed) under the condition of constant concentration in the compartment with the inward transmembrane volume flow (permeate), compartments (') and (") in Fig3, respectively. The latter of the abovementioned conditions means that $\left(J^{ch}\right)'' = 0$. Consequently, the definition of the rejection, $\gamma$, takes the form

$$\gamma = -\left[\frac{\left(J^{ch}\right)'}{J_v^m c}\right]_{\substack{\left(J^{ch}\right)''=0 \\ I=0}} \tag{59}$$

Also, we introduce the reflection coefficients of individual ions, $\sigma_A$ and $\sigma_B$, that are defined with help of a thought experiment where the electric potentials of the adjacent solutions are maintained to be equal

$$\sigma_{A,B} = -\left[\frac{\left(J_{A,B}^{ch}\right)^m}{J_v^m c_{A,B}}\right]_{\substack{\Delta c=0 \\ \Delta\Phi=0}} = \left(1 - \frac{J_{A,B}^m}{J_v^m c_{A,B}}\right)_{\substack{\Delta c=0 \\ \Delta\Phi=0}} \tag{60}$$

Note, that, in the thought experiment considered in definition (60), the transmembrane current is not zero and the VEC in Fig.3 works with the electrodes shortly connected via external circuit

*5.3 Electrochemical coupling. Transport and transference numbers*

The parameters responsible for selective transport of different ions through a membrane while passing the transmembrane electric current are referred to as the transport numbers. In the literature [1, 2, 27, 29-31, 35-46, 83, 84], one can face at least four types of transport numbers. All of them can be obtained from the results of measuring the rate of composition changes produced by the electric current passed through the membrane at equal concentrations in the adjacent solutions.

To define the first two types of the transport numbers, we consider the overall rate of concentration changes in the cell shown in Fig.3. As stated in Section 5.1, such a rate is given by the ion chemical fluxes, $\left(J_{A,B}\right)^{ch}$, that are common for both the solutions and defined by Eq.(53). According to Eq.(55), the chemical flux is split in the transmembrane and electrode parts, $\left(J_{A,B}^{ch}\right)^m$ and $\left(J_{A,B}^{ch}\right)^e$, respectively. When $\left(J_{A,B}\right)^{ch}$ have been measured and $\left(J_{A,B}^{ch}\right)^e$ are preliminary known from the data on electrode properties, one can determine $\left(J_{A,B}^{ch}\right)^m$. The obtained $\left(J_{A,B}^{ch}\right)^m$ can be used for obtaining the ion transport numbers attributed to the membrane, $t_{A,B}$. The abovementioned two types of the transport numbers differ from each other by the hydraulic regime of measurements, $J_v = 0$ and $\Delta p = 0$

$$t_{A,B} = \left[\frac{\left(J^{ch}_{A,B}\right)^m Fz_{A,B}}{I}\right]_{\substack{\Delta c=0 \\ J^m_v=0}} = \left[\frac{J^m_{A,B} Fz_{A,B}}{I}\right]_{\substack{\Delta c=0 \\ J^m_v=0}} \quad \text{(a)}$$

$$t^*_{A,B} = \left[\frac{\left(J^{ch}_{A,B}\right)^m Fz_{A,B}}{I}\right]_{\substack{\Delta c=0 \\ \Delta p=0}} = \left[\frac{J^m_{A,B} Fz_{A,B}}{I}\right]_{\substack{\Delta c=0 \\ \Delta p=0}} - Fz_{A,B} c_{A,B} \left(\frac{J^m_v}{I}\right)_{\substack{\Delta c=0 \\ \Delta p=0}} \quad \text{(b)}$$
(61)

The latter equality in Eq.(61a) follows from Eq.(56a). Clearly,

$$t_A + t_B = t^*_A + t^*_B = 1 \tag{62}$$

Thus, for obtaining the ion transport numbers of two types discussed above, $t_{A,B}$ and $t^*_{A,B}$, one should measure the overall rates of the concentration changes under the respective hydraulic conditions indicated in Eqs.(61a) and (61b). Such measurements yields the chemical flux, $J^{ch}_{A,B}$, given by Eqs.(51) and (53) for these conditions. If one knows as well electrode chemical fluxes $\left(J^{ch}_{A,B}\right)^e$ defined by Eq.(56b), the use of Eq.(55) yields $\left(J^{ch}_{A,B}\right)^m$ for the respective hydraulic conditions. Consequently, using Eqs.(61a) and (61b) enables obtaining $t_{A,B}$ and $t^*_{A,B}$.

A reasonable approach in executing the scheme outlined above is to use an electrode couple blocking the electrochemical reaction for the ion whose transport number is determined. Remarkably, in such a case, the electrode chemical flux of this ion is not zero because, in the presence of electric current, there is an electrode volume flow, $J^e_v$. The latter flow is not zero since the other ion, which is not blocked, definitely takes part in the electrode reactions thereby resulting in the electrode contribution into changing the solution volume.

To illustrate the above statement, let us consider an electrode blocking the ion $A$ or $B$. According to Eq.(56b), the chemical flux of ion $A$ or $B$ toward the electrode blocking the ion $A$ or $B$, $\left(J^{ch}_A\right)^{e(A)}$ or $\left(J^{ch}_B\right)^{e(B)}$, are given, respectively, as

$$\left(J^{ch}_A\right)^{e(A)} = -c_A J^{e(A)}_v = -\frac{I}{Fz_B} v_A c v_B$$

$$\left(J^{ch}_B\right)^{e(B)} = -c_B J^{e(B)}_v = -\frac{I}{Fz_A} v_B c v_A$$
(63)

While using Eqs.(55) and (63), the transmembrane chemical fluxes of ions $A$ and $B$, $\left(J^{ch}_A\right)^m$ and $\left(J^{ch}_B\right)^m$, are expressed in the following form

$$\left(J_A^{ch}\right)^m = \left(J_A^{ch}\right)^{(A)} - \frac{I}{Fz_B}v_A cv_B$$
$$\left(J_B^{ch}\right)^m = \left(J_B^{ch}\right)^{(B)} - \frac{I}{Fz_A}v_B cv_A \qquad (64)$$

Consequently, applying the definitions given by Eqs.(61) and using the electroneutrality condition given by Eq.(5) yield

$$t_{A,B} = (1-cv)t_{A,B}^H + cv_{B,A}v_{B,A}$$
$$t_{A,B}^* = (1-cv)\left(t_{A,B}^*\right)^H + cv_{B,A}v_{B,A} \qquad (65)$$

In Eq.(65), we introduced the Hittorf transport numbers, $t_{A,B}^H$ and $\left(t_{A,B}^*\right)^H$ that are named after the method inventor [83,84]. They are two other types of the transport numbers. Obtaining these parameters is linked to the equations

$$t_{A,B}^H = \left(\frac{Fz_{A,B}v_{A,B}J_{A,B}}{I}\right)_{\substack{\Delta C=0 \\ J_v^m=0}}$$
$$\left(t_{A,B}^*\right)^H = \left(\frac{Fz_{A,B}v_{A,B}J_{A,B}}{I}\right)_{\substack{\Delta C=0 \\ \Delta p=0}} \qquad (66)$$

where we introduced the flux $J_{A,B}$ which is expressed with the help of Eqs.(7), (56a), (64) and (65) and some transformations, as

$$J_{A,B} = \frac{\left(J_{A,B}^{ch}\right)^{(A)}}{1-cv} = \left(M_w'\frac{dm}{d\tau}\right)^{(A,B)} = \left(M_w''\frac{dm}{d\tau}\right)^{(A,B)} \qquad (67)$$

In Eq.(67), the superscripts signify the ion which is blocked by the employed electrodes; Note that the second and third expressions in each of Eqs.(67) look somewhat similarly to the definition of the chemical flux given by Eq. (50). However, the solution volume, $V$, and concentration $c$ in the respective compartments are replaced in Eqs.(67) by, respectively, the electrolyte solvent mass $M_w$ and molality, $m = n/M_w$.

The Hittorf transport numbers defined by Eqs.(66) have some remarkable properties. With the help of Eqs.(38), (62) and (65), one can see that

$$t_A^H + t_B^H = \left(t_A^*\right)^H + \left(t_B^*\right)^H = 1 \qquad (68)$$

For sufficiently diluted solutions, $cv \to 0$, the Hittorf transport numbers approach to the transport numbers defined by Eqs. (61), $t_{A,B}^H \to t_{A,B}$; $\left(t_{A,B}^*\right)^H \to t_{A,B}^*$; $t_A^H + t_B^H \to 1$ and $\left(t_A^*\right)^H + \left(t_B^*\right)^H \to 1$. As stated in Section 3.2, the approximation $c_{A,B}|v_{A,B}| \ll 1$ is employed in the KKM theory [30, 31]. Hence, within the frameworks of the KKM approach, the transport numbers given by Eq.(61) can be considered as the Hittorf transport numbers.

Additionally, one can introduce another group of parameters that will be referred to as the ion transference numbers, $\theta_{A,B}$ and $\theta_{A,B}^*$. These coefficients characterize coupling between the electric current and the transmembrane ion fluxes at zero volume flow and pressure difference, respectively,

$$\theta_{A,B} = \left(\frac{J_{A,B}^m F}{I}\right)_{\substack{\Delta c=0 \\ J_v^m=0}} = \frac{t_{A,B}}{z_{A,B}} \quad \text{(a)}$$

$$\theta_{A,B}^* = \left[\frac{\left(J_{A,B}^{ch}\right)^m F}{I}\right]_{\substack{\Delta c=0 \\ \Delta p=0}} = \left(\frac{J_{A,B}^m F}{I}\right)_{\substack{\Delta c=0 \\ \Delta p=0}} - Fc_{A,B}\left(\frac{J_v^m}{I}\right)_{\substack{\Delta c=0 \\ \Delta p=0}} = \frac{t_{A,B}^*}{z_{A,B}} \quad \text{(b)}$$

(69)

The second equality in Eq.(69a) directly follows from Eq.(61a) whereas the latter equality in Eq.(69b) is obtained by combining Eqs.(56a) and (61b).

## 5.4 Electromechanical (Electrokinetic) coupling

Now, we focus on two Electrokinetic Phenomena being manifestations of electromechanical coupling of fluxes. The Streaming Current, which was first mentioned in the list given in the end of Section 1.3, amounts to generating the transmembrane electric current by passing through the membrane a volume flow at zero electric potential difference across the membrane. We will describe this coupling with the help electrokinetic charge density, $\rho\left(C/m^3\right)$, defined as

$$\rho = \left(\frac{I}{J_v^m}\right)_{\substack{\Delta c=0 \\ \Delta\Phi=0}} \tag{70}$$

Realize that the conditions indicated above for measuring $\rho$ are exactly the same as those presented in Eq.(60), which defines the individual ion reflection coefficients, $\sigma_A$ and $\sigma_B$. Consequently, by combing Eqs.(6), (7), (60) and (70), one obtains

$$\rho = Fcv_A z_A (\sigma_B - \sigma_A) = Fcv_B z_B (\sigma_A - \sigma_B) \tag{71}$$

Inspecting Eq.(71) reveals that the sign of $\rho$ coincides with the sign of charge of the ion whose reflection coefficient is lesser.

Electroosmosis is the second Electrokinetic Phenomenon mentioned in the Section 1.3. It is observed when the transmembrane electric current is passed through the membrane and produces the transmembrane volume flow due to the electromechanical coupling of fluxes at zero pressure difference across the membrane. For characterizing this coupling, it is convenient to use the electroosmotic coefficient, $K_{eo}\left(m^3/C\right)$.

$$K_{eo} = \left(\frac{J_v^m}{I}\right)_{\substack{\Delta c = 0 \\ \Delta p = 0}} \quad (72)$$

Although $\rho$ and $K_{eo}$ have mutually inverse dimensions, they are interrelated by a more complex relationship than the inverse proportionality. This relationship cannot be derived within the framework of purely kinematic analysis considered here.

While using the definition given by Eq.(72), one can rewrite Eq.(69b) as

$$\theta_{A,B}^* = \left(\frac{J_{A,B}^m F}{I}\right)_{\substack{\Delta c = 0 \\ \Delta p = 0}} - F c_{A,B} K_{eo} \quad (73)$$

In the literature, one can face the following coefficient describing the solvent transfer in the presence of electric current

$$\theta_w = \left(\frac{J_w F}{I}\right)_{\substack{\Delta c = 0 \\ J_v = 0}} \quad (a)$$

$$\theta_w^* = \left(\frac{J_w F}{I}\right)_{\substack{\Delta c = 0 \\ \Delta p = 0}} \quad (b) \quad (74)$$

According to the analogy with the coefficient defined by Eqs.(69), we name $\theta_w$ and $\theta_w^*$ as the transference numbers of solvent attributed to the respective hydraulic regime of measurements. Note that, in some publications (see refs. [1, 38, 39], for example), the quantity $\theta_w^*$ is referred to as the transport number of solvent.

By combining Eqs.(7), (37) and (38) with the definitions given by Eqs.(69) and (72)-(74), the solvent transference numbers, $\theta_w$ and $\theta_w^*$, are interrelated with the ion transference numbers, $\theta_{A,B}$ and $\theta_{A,B}^*$, and the electroosmotic coefficient, $K_{eo}$, as

$$\theta_w = -\frac{\theta_A v_A + \theta_B v_B}{v_w} \quad (a)$$

$$\theta_w^* = \frac{F K_{eo}(1 - cv) - \theta_A^* v_A - \theta_B^* v_B}{v_w} \quad (b) \quad (75)$$

Remarkably, Eq.(75b) yields a relationship between the parameters responsible for electrochemical and electromechanical transformations, respectively, $\theta_w^*$; $\theta_{A,B}^*$ and $K_{eo}$. Next, we obtain a number of relationships that interconnect coefficients responsible for different type of free energy transduction.

*5.5 Relationships between some coefficients describing coupling of fluxes*

In the present Section we will obtain several important relationships that interrelate the coefficient responsible for membrane selectivity and those responsible for Electrokinetic Phenomena.

In the simultaneous presence of the transmembrane electric current and volume flow, for equal concentration in the adjacent compartments, the transmembrane ion flux can be represented as the following linear superposition.

$$J_{A,B}^m = I l_{A,B}^I + J_v^m l_{A,B}^{J_v^m} \tag{76}$$

The above equation can be considered as a the Taylor expansion of the function of two variables, $J_{A,B}^m(I, J_v^m)$, where the leading, linear, terms are retained, only. Note that the zero order term is definitely equal to zero.

In the linear case, the coefficients $l_{A,B}^I$ and $l_{A,B}^{J_v^m}$ in Eq.(76) are independent of the transmembrane electric current and volume flow and can be determined by analyzing particular regimes. Consequently, by using simultaneously Eqs.(58), (61a) and (76), one obtains

$$J_{A,B}^m = \frac{I t_{A,B}}{F z_{A,B}} + J_v^m v_{A,B} c(1-\sigma) \tag{77}$$

Let us consider now the regime of measuring Streaming Current. The respective conditions are indicated in Eqs. (70). Accordingly, we replace $J_A^m$ and $I$ in Eq. (77) with the help of Eqs. (60) and (70) to obtain

$$(1-\sigma_{A,B}) c_{A,B} \left( J_v^m \right)_{\substack{\Delta c=0 \\ \Delta\Phi=0}} = \frac{\rho}{F z_{A,B}} t_{A,B} \left( J_v^m \right)_{\substack{\Delta c=0 \\ \Delta\Phi=0}} + (1-\sigma) c_{A,B} \left( J_v^m \right)_{\substack{\Delta c=0 \\ \Delta\Phi=0}} \tag{78}$$

Obvious transformation of Eq.(78) conducted with the help of Eq.(71) yields

$$\sigma = \sigma_A t_B + \sigma_B t_A \tag{79}$$

The above result, which interrelates the transport numbers defined by Eq.(61a) and the reflection coefficients defined by Eqs.(58) and (60), coincides with that derived in ref. [35]. Note that the above derivation was conducted by using the purely kinematic analysis.

Now, we will derive an important relationship between two types of transport numbers defined by Eqs (61).To this end, Eq. (61b) should be rewritten, as

$$t_A^* = \left[ \frac{\left( J_A^m - c_A J_v^m \right) F z_A}{I} \right]_{\substack{\Delta c=0 \\ \Delta p=0}} \tag{80}$$

The next step in the derivation is to adapt the general relationship given by Eq.(77) to the regime indicated in Eq.(61b).

$$\left(J_A^m\right)_{\substack{\Delta c=0\\\Delta p=0}} = I_{\substack{\Delta c=0\\\Delta p=0}}\frac{t_A}{Fz_A} + \left(J_v^m\right)_{\substack{\Delta c=0\\\Delta p=0}}\nu_A c(1-\sigma) \tag{81}$$

After a short transformation, combining Eqs.(72), (80) and (81), yields

$$t_{A,B}^* = t_{A,B} - Fc\nu_{A,B}z_{A,B}\sigma K_{eo} \tag{82}$$

Thus, Eq.(82) gives a relationship between the transport numbers measured at zero volume flow, $t_{A,B}$, and those measured at zero pressure difference, $t_{A,B}^*$.

Combining Eqs.(71), (79) and (82) enables us to express the electrolyte reflection coefficient through the transport numbers measured at zero pressure difference, $t_{A,B}^*$

$$\sigma = \frac{\sigma_A t_A^* + \sigma_B t_B^*}{1 + K_{eo}\rho} \tag{83}$$

Now, we express the electroosmotic coefficient, $K_{eo}$, in terms of transference numbers $\theta_{A,B}^*$ and $\theta_w^*$ defined by Eqs.(69b) and (75b) for the regime of $\Delta p = 0$. While resolving Eq.(75b) with respect to $K_{eo}$ one obtains

$$K_{eo} = \frac{1}{F}\left(\theta_w^* \mathrm{v}_w + \theta_A^* \mathrm{v}_A + \theta_B^* \mathrm{v}_B\right) \tag{84}$$

By eliminating in Eq.(84) either $\mathrm{v}_A$ or $\mathrm{v}_B$ with the help of Eq. (37) and using Eqs.(5), (61) and (69), one rewrites Eq.(84) in two following equivalent forms

$$K_{eo} = \frac{1}{F}\left(\theta_w^* \mathrm{v}_w + \frac{\theta_A^*}{\nu_A}\mathrm{v} + \frac{\mathrm{v}_B}{z_B}\right) = \frac{1}{F}\left(\theta_w^* \mathrm{v}_w + \frac{\theta_B^*}{\nu_B}\mathrm{v} + \frac{\mathrm{v}_A}{z_A}\right) \tag{85}$$

In the brackets of any of two equivalent results given by Eq.(85), while taking into account the first two terms, only, we obtain an expression which is completely equivalent to that derived on the basis of the KKM theory in ref.[38] (Eq.(7) of the reference). The simultaneous existence of two different correct versions corresponds to the dualism of the KKM theory which was discussed in details in Section 3.2. The third term in the brackets in both the expressions of Eq.(85) enables a generalization of the result reported in ref.[38] by accounting for the term missed in the KKM expression for the transmembrane volume flow. Recall Eq. (42) derived in Section 4.2 where each version of the KKM volume flows, $J_v^{(A,B)}$, is corrected by an additional term. Note that additions of the third terms in the expressions of Eq.(85) make the results equal whereas accounting for two terms, only, yields different predictions for $K_{eo}$.

Importantly, the correcting term in Eq.(85), $\mathrm{v}_{B,A}/z_{B,A}$, has the same order of absolute value as the second term in the brackets, $\theta_{A,B}^*\mathrm{v}/\nu_{A,B}$, and, in many cases, exceeds it. The latter can be illustrated by example of $MgCl_2$ for which $\mathrm{v}/\nu_{Cl^-} \approx 7.6\ cm^3/mol$ whereas $v_{Mg^{2+}}/z_{Mg^{2+}} \approx -16 \ cm^3/mol$ (see the data listed in Sections 3.1 and 4.3). Hence, the corrections presented by the third term in brackets of Eq.(85) is important at least when the second term, which is taken into account in ref [38], yields a noticeable contribution

**6. Conclusions**

1) At none-zero transmembrane electric current, an analysis of transmembrane fluxes and changes in the composition of adjacent solutions should unavoidably consider a hypothetical or real couple of electrodes being a source and sink of the electric current. In the presence of such electrodes, the changes in the adjacent solution volumes and compositions, additionally to membrane, depend on the electrode properties. In the paper, we refer to such an electrode system as the Virtual Electrochemical Cell (VEC)

2) In the Kedem - Katchalsky – Michaeli (KKM) theory, the VEC is formed by two identical electrodes that block discharge of one of the ions and are ideally reversible with respect to other one. Clearly, for any binary electrolyte, there are two versions of the KKM VEC.

3) Two of three thermodynamic fluxes chosen in the KKM theory are the rates of changing the solution volume and electrolyte amount in the VEC compartments. However, in the KKM theory, these fluxes are identified with the transmembrane volume flow and electrolyte flux, respectively. In the present paper, it is stressed that such identifications are correct for zero electric current, only. For non-zero electric current, the aforementioned identifications can serve as an approximation which gives a good description for sufficiently low electrolyte concentrations.

4) An important advantage of the abovementioned KKM approximation is that, for sufficiently low electrolyte concentrations, it gives a description in purely thermodynamic terms, i.e., using no extra-thermodynamic assumptions. Such an advantage is a result of using the partial molar volume of electrolyte while defining the transmembrane volume flow. The partial molar volume of electrolyte is a directly measured equilibrium thermodynamic parameter of the adjacent solutions. The latter is clear from the short review of the literature dealing with the widely used complex method of experimental obtaining this quantity. The method is based on measuring the concentration dependency of solution density and certain mathematical scheme of extracting the partial molar volume of solute from these dependencies.

5) At non-zero current, the KKM approximation leads to an appreciable error when the electrolyte concentration is sufficiently high. In such situations, it is unavoidable to define the transmembrane volume flow by using such parameters as the partial molar volumes of individual ions. In the studies reported in the literature, these parameters are obtained with the help of sophisticated scheme based on measuring partial molar volume for series of binary electrolytes with a common ion, using the Yung additivity rule and making certain extra-thermodynamic assumptions.

6) A big massive of experimental and theoretical literature on the partial molar volume of inorganic ions reveals a number of remarkable regularities. The most important of them is the relatively large partial molar volume for anions which can reach the value of about $50 \ cm^3/mol$ and negative value of that for multi-charged cations of about $-50 \ cm^3/mol$, in some cases. The latter is a manifestation of electrostriction produced by the ion. As stated in the reviewed literature the difference in the behavior of cations and anions is explained through the differences in their intrinsic volumes.

7) An important flux introduced in the paper is referred to as the chemical flux which is equal to the overall rate of concentration changes in each of the compartments separated by the membrane. In the linear case, the chemical fluxes attributed to different compartments turn out to be equal. This quantity is convenient for defining the kinetic coefficients responsible for membrane selectivity in different transfer processes. It should be measured in the experiments with mechano-chemical coupling of fluxes that are intended for obtaining refection coefficients of solute and individual ions. As well, measuring the chemical fluxes enables one to obtain different types of the ion transport numbers in the experiments with the electrochemical coupling.

8) A set of fundamental relationships has been derived to interrelate the kinetic coefficients describing Electrokinetic Phenomena and those characterizing selectivity of membrane transfer in different hydraulic and galvanic regimes. The relationships are obtained while conducting the kinematic analysis of fluxes that are produced by the simultaneously imposed transmembrane electric current and volume flow. Remarkably, the derivation does not deal with the Thermodynamic Forces and does not use the Onsager Theorem.

9) The obtained results can be used in

- extending the KKM theory for the case of arbitrary concentrated electrolyte solutions;
- addressing the electrochemical membrane separation and energy conversion effects as well as the respective technological schemes for the case of electrolyte solution having concentrations close to $10^3 \ mol/m^3$ or more;
- interpreting data of electrokinetic experiments conducted in solutions with the abovementioned high electrolyte concentrations where the contribution of ion migration into the volume flow can be appreciable;

- characterizing the selective properties of a membrane in concentrated solutions with the help of electrochemical and electrokinetic measurements;
- mathematical modeling the processes of electrokinetic/electrodialytic remediation of soils and clays by removing pollutants containing heavy and/or organic ions;
- exploring the possibilities of measuring the partial molar volumes of individual ions by combining electrochemical and electrokinetic measurements

**List of Symbols**

*Latin Letters*

$A, B$ - notation of ion

$c_{A,B}$ - molar concentrations of ions

$c_w$ - molar concentration of solvent

$c$ - molar concentration of electrolyte

$d$ - mass density of solution

$d_w$ - mass density of pure solvent

$E$ - "electromotive force" in the KKM theory

$F$ - Faraday constant

$I$ - electric current

$J_{A,B}^m$ - transmembrane ion fluxes

$J_{A,B}^*$ - apparent ion fluxes

$J_{A,B}^e$ - electrode flux of ion

$J_w^m$ - transmembrane solvent flux

$J_v^m$ - transmembrane volume flow

$J_v^*$ - apparent volume flow

$J^*$ - apparent electrolyte flux

$J_s$ - electrolyte flux in the KKM theory

$J^{ch}$ - chemical flux of electrolyte

$J_{A,B}^{ch}$ - chemical flux of ion

$\left(J_{A,B}^{ch}\right)^m$ - transmembrane chemical flux of ion

$\left(J_{A,B}^{ch}\right)^e$ - electrode chemical flux of ion

$K_{eo}$ - electroosmotic coefficient

**L** - matrix of kinetic coefficients

M - molar mass of electrolyte

m - molality

N - Avogadro number

$n$ - electrolyte amount

$n_{A,B}$ - amount of ion

$p$ - pressure

$s$ - entropy

$T$ - absolute temperature

$t_{A,B}$ - ion transport numbers at zero transmembrane volume flow

$t_{A,B}^*$ - ion transport numbers at zero pressure difference

$t_{A,B}^H$ - Hittorf transport numbers of ions at zero transmembrane volume flow

$\left(t_{A,B}^H\right)^*$ - Hittorf transport numbers of ions at zero pressure difference

$V$ - volume of compartment

$v_{A,B}$ - molar volumes of ions

$v_w$ - molar volume of solvent

$v$ - electrolyte molar volume

$v^*$ - apparent molar volume of electrolyte

$W$ - Entropy Production Function

**X** - vector of thermodynamic forces

**Y** - vector of thermodynamic fluxes

$z_{A,B}$ - ionic charges

*Greek letters*

$\gamma$ - rejection

$\Delta$ - difference between intensive thermodynamic parameters

$\theta_{A,B}$ - ion transference numbers at zero transmembrane volume flow

$\theta_{A,B}^*$ - ion transference numbers at zero pressure difference

$\theta_w$ - transference number for solvent at zero transmembrane volume flow

$\theta_w^*$ - transference number for solvent at zero pressure difference

$\Phi$ - electric potential

$\mu_k$ - chemical potentials of the kth solution component

$\nu_{A,B}$ - stoichiometric coefficients for reaction of electrolyte dissociation

$\Pi$ - osmotic pressure

$\rho$ - electrokinetic charge density

$\sigma$ - reflection coefficient of electrolyte

$\sigma_{A,B}$ - reflection coefficient of ion

**Acknowledgments**

We gratefully acknowledge the financial support by European Commission within the scope of FP Horizon 2020 (Project acronym "EHAWEDRY," Grant Agreement No. 964524) and National Research Foundation of Ukraine (project number 2020.02/0138).

**Appendix: The partial and apparent molar volumes. Derivation of major relationships**

We consider the changes of volume by mixing the mass $M_w$ of a solvent having volume $V_w$ and mass density $d_w = M_w/V_w$ with $n$ moles of a solute having molar mass M and mass $M - M_w = nM$. In the latter equality, $M$ is the mass of the obtained solution having the volume $V$; the mass density $d = M/V$; the molar solute concentration per unity of solution volume $c = nd/M$ and the molality $m = n/M_w$.

*A1. Derivation of phenomenological relationships for obtaining the apparent and partial molar volumes of electrolyte*

Consequently, in order to express the apparent molar volume in terms of the measured densities d and $d_w$, one can suggest the following identical transformations of the definition given by the first equality of Eq.(21)

$$v^* = \left(\frac{V-V_w}{n}\right)_{p,T,N_w} = M\frac{\frac{M}{d}-\frac{M_w}{d_w}}{M-M_w} = M\frac{\frac{M}{d}-\frac{M}{d_w}-\frac{M_w-M}{d_w}}{M-M_w} = \frac{M}{d_w}+M\frac{\frac{M}{d}-\frac{M}{d_w}}{M-M_w} = \frac{M}{d_w}+M\frac{M}{dd_w}\frac{d_w-d}{M-M_w} = \frac{M}{d_w}+M\frac{V}{M-M_w}\frac{d_w-d}{d_w} = \frac{M}{d_w}-\frac{d-d_w}{cd_w} \quad (A1)$$

or

$$v^* = \left(\frac{V-V_w}{n}\right)_{p,T,n_w} = M\frac{\frac{M}{d} - \frac{M_w}{d_w}}{M - M_w} = M\frac{\frac{M-M_w}{d} + \frac{M_w}{d} - \frac{M_w}{d_w}}{M - M_w} = \frac{M}{d} + \quad (A2)$$

$$M\frac{M_w}{M-M_w}\frac{d_w - d}{dd_w} = \frac{M}{d} - \frac{d - d_w}{mdd_w}$$

Thus, the final expressions of Eqs.(A1) and (A2) give the results presented in Eq.(25)

The definition given by the first equality of Eq.(21) yields

$$V = nv^* + V_w \qquad (A3)$$

By combining Eqs.(25) and (A3) one obtains

$$v = v^* + n\left(\frac{\partial v_*}{\partial n}\right)_{p,T,n_w} = v^* + m\left(\frac{\partial v_*}{\partial m}\right)_{p,T,n_w} \qquad (A4)$$

Using the definition of molality and Eq.(A3) gives

$$m = \frac{n}{M_w} = \frac{cV}{d_w V_w} = \frac{c(nv^* + V_w)}{d_w V_w} = c\left(mv^* + \frac{1}{d_w}\right) \qquad (A5)$$

Consequently, the molality is expressed as

$$m = \frac{c}{d_w(1-cv_*)} \qquad (A6)$$

Differentiating both sides of Eq.(A6) enables one to obtain

$$dm = \frac{1 + c^2\frac{\partial v_*}{\partial c}}{(1-cv_*)^2 d_w} dc \qquad (A7)$$

While combining Eqs.(A4) and (A7), we arrive at the relation expressing the partial molar volume, v, through the function $v_*(c)$ and its derivative

$$v = v^* + m\frac{\partial v_*}{\partial m} = v^* + \frac{1-cv^*}{1+c^2\frac{\partial v^*}{\partial c}} c\frac{\partial v^*}{\partial c} \qquad (A8)$$

Thus, Eq.(A8) coincides with Eq.(26).

*A2. Derivation of expressions describing impact of the electrostriction effect on the partial ion volumes of ions and electrolytes*

*A.2.1 Local density changes and partial molar volume of ion*

Now, we consider how the reorganization of solvent by the added solute changes the apparent molar volume. To this end, we introduce the local solvent density around a given ion, A or B, $d_l^{A(B)}(\mathbf{r})$, which is a function of the coordinate, $\mathbf{r}$. For the case of low electrolyte concentration, we assume that the changes of density produced by the individual ions are additive. It is convenient to introduce the local perturbation of density due to the added ion $A$ (or $B$), $\delta d^{A(B)}(\mathbf{r})$

$$\delta d^{A(B)}(\mathbf{r}) = d_l^{A(B)}(\mathbf{r}) - d_w \tag{A9}$$

Consequently, the change of solvent density produced by $n_{A(B)}$ moles of the added ion $A$ (or $B$) is given by

$$d^{A(B)} - d_w = \frac{n_{A(B)} N}{V} \int_V \delta d^{A(B)}(\mathbf{r}) dV = c_{A(B)} N \int_V \delta d^{A(B)}(\mathbf{r}) dV = c\nu_{A(B)} N \int_V \delta d^{A(B)}(\mathbf{r}) dV \tag{A10}$$

where N is the Avogadro number; $c_{A(B)} = n_{A(B)}/V = c\nu_{A(B)}$ is the molar concentration of respective ion.

We divide the difference between the solution and pure solvent densities, $d - d_w$, into two parts, namely, the solute contribution, $Mc$, and the change due to the solvent reorganization, $\Delta d_{reorg}$, as

$$d - d_w = Mc + \Delta d_{reorg} \tag{A11}$$

With the help of Eq.(A10), the solvent reorganization term, $\Delta d_{reorg}$, is represented in the form

$$\Delta d_{reorg} = cN \int_V \left[ \nu_A \delta d^A(\mathbf{r}) + \nu_B \delta d^B(\mathbf{r}) \right] dV = -c\mathrm{v}_{int} d_w + cN \int_{V-V_{solute}} \left[ \nu_A \delta d^A(\mathbf{r}) + \nu_B \delta d^B(\mathbf{r}) \right] dV \tag{A12}$$

where $\mathrm{v}_{int}$ is the solute intrinsic partial molar volume which coincides with the intrinsic apparent molar volume and is given by

$$\mathrm{v}_{int} = NV_{int}^{solute} = N\left(\nu_A V_{int}^A + \nu_B V_{int}^B\right) \tag{A13}$$

In Eq.(A13), $V_{int}^{solute}$ is the volume occupied by the ions originating from the dissociation of one electrolyte molecule according to Eq.(4). Consequently, the last integration in Eq.(A12) is conducted over the whole volume outside the representative ions.

By combining the last expression of Eq.(A1) with Eqs.(A11)-(A13), we arrive at the following common expression for the partial and apparent molar volumes of electrolyte

$$\mathrm{v} = \mathrm{v}_* = \nu_A \mathrm{v}^A + \nu_B \mathrm{v}^B = \nu_A \mathrm{v}_*^A + \nu_B \mathrm{v}_*^B = N\left(\nu_A V_{int}^A + \nu_B V_{int}^B\right) + \nu_A \mathrm{v}_{reorg}^A + \nu_B \mathrm{v}_{reorg}^B \tag{A14}$$

where the contributions into the ion apparent molar volumes due to the medium reorganization, $v^A_{reorg}$ and $v^B_{reorg}$ take the forms

$$v^{A,B}_{reorg} = -\frac{N}{d_w} \int_{V-V_A} \delta d^{A,B}(\mathbf{r}) dV \tag{A15}$$

Thus, for the known intrinsic volumes, $V^{solute}_{int}$, $V^A_{int}$, $V^B_{int}$ and mass density perturbation around each of the ions, $\delta d^A(\mathbf{r})$ and $\delta d^B(\mathbf{r})$, by using Eqs.(A14) and (A15), one can predict the apparent molar volumes for both the electrolyte and the ions.

*A2.2 Derivation of the Drude-Nernst equation for the electrostriction impact on partial molar volume*

Let us consider an individual ion having charge $z$ in a solvent which is assumed to be a perfect dielectric whose permittivity depends on the mass density d and thus on the pressure $p$, $\varepsilon_w = \varepsilon_w[d_w(p)]$. For such a liquid surrounding, the hydrostatic momentum balance is written in the form

$$-\nabla p + \frac{1}{2}\nabla\left[E^2 d_w\left(\frac{\partial \varepsilon_w}{\partial d_w}\right)_T\right] - \frac{1}{2}E^2 \nabla \varepsilon_w = 0 \tag{A16}$$

where $E$ is the magnitude of the electric field strength, $\mathbf{E}(\mathbf{r})$, produced by the ion under consideration. In Eq.(A16), the first term yields the mechanic force whereas two other terms yield the Helmholtz electric force consisting of the electrostriction part and the part proportional to the permittivity gradient, respectively [81].

At constant temperature, $d = d(p)$ and $\varepsilon = \varepsilon[d(p)]$. Consequently, Eq.(A16) is rewritten, as

$$-\nabla d + \frac{1}{2}\frac{\partial d_w}{\partial p} d \nabla\left[E^2\left(\frac{\partial \varepsilon_w}{\partial d_w}\right)_T\right] = 0 \tag{A17}$$

Now, we represent the unknown function $d(\mathbf{r})$ in the form given by Eq.(A9) and substitute it in Eq.(A17)

$$-\nabla \delta d + \frac{1}{2}d_w \nabla\left[E^2\left(\frac{\partial \varepsilon_w}{\partial p}\right)_T\right] + \delta d \frac{\partial d_w}{\partial p}\nabla\left[E^2\left(\frac{\partial \varepsilon_w}{\partial d_w}\right)_T\right] = 0 \tag{A18}$$

By taking into account that $\nabla \cdot \varepsilon \mathbf{E} = 0$ outside the ion and assuming the spherical symmetry, one obtains.

$$E = \frac{ze}{4\pi\varepsilon(r)r^2} = \frac{ze}{4\pi r^2 \varepsilon_w}\left[1 - \frac{\delta\varepsilon(r)}{\varepsilon_w}\right] \tag{A19}$$

where $\delta\varepsilon(r) = \varepsilon(r) - \varepsilon_w$. While realizing that $\delta\varepsilon(r)/\varepsilon_w == O(E^2 \partial\varepsilon/\partial p)$ and $\delta d/d_w = O(E^2 \partial\varepsilon_w/\partial p)$, combining Eqs.(18) and (19) and retaining the first order perturbation terms, only, we obtain

$$-\nabla\left[\delta d - \frac{1}{2}d_w\left(\frac{ze}{4\pi r^2 \varepsilon_w}\right)^2 \left(\frac{\partial\varepsilon_w}{\partial p}\right)_T\right] = 0 \tag{A20}$$

Since $\delta d \to 0$ when $r \to \infty$, Eq.(A20) is easily integrated to yield the mass density perturbation, $\delta d(r)$

$$\frac{\delta d}{d_w} = \frac{\delta d_{D-N}}{d_w} = \frac{1}{2\varepsilon_w}\left[\frac{\partial\ln(\varepsilon_w)}{\partial p}\right]_T \left(\frac{ze}{4\pi r^2}\right)^2 \tag{A21}$$

where we introduced the notation $\delta d_{D-N}$ intended to distinguish the Drude-Nernst mass density perturbation from that predicted by Redlich and Rosenfeld [50] by using the Debye-Hückel model [79]. Any of Eqs.(15) can be rewritten as

$$v_{reorg} = -4\pi N \int_a^\infty \frac{\delta d_{D-N}(r)}{d_w} r^2 dr \tag{A22}$$

By combing Eqs.(A21) and (A22), we obtain the molar volume due to the reorganization of solvent, $v_{reorg}$, in the form

$$\begin{aligned} v_{reorg}^A &= -\frac{N}{8\pi\varepsilon_r\varepsilon_0}\left[\frac{\partial\ln(\varepsilon_r)}{\partial p}\right]_T \frac{z_A^2 e^2}{a} \\ v_{reorg}^B &= -\frac{N}{8\pi\varepsilon_r\varepsilon_0}\left[\frac{\partial\ln(\varepsilon_r)}{\partial p}\right]_T \frac{z_B^2 e^2}{b} \end{aligned} \tag{A23}$$

where $a$ and $b$ are the ionic radii; $\varepsilon_0$ is the universal dielectric constant; $\varepsilon_r = \varepsilon_w/\varepsilon_0$ is the relative permittivity of the solvent.

Thus, any of Eqs.(A23) gives the Drude-Nernst result for affect of electrostriction on the partial molar volume of ions. Within the frameworks of the Drude-Nernst model, the total partial molar volume of ion is a sum of intrinsic and electrostriction parts. Since the partial molar and apparent volumes of electrolyte turn out to be equal because of the independency of electrolyte concentration, $v_* = v$, we obtain

$$\begin{aligned} v^A &= v_*^A = \frac{4}{3}\pi N a^3 - \frac{N}{8\pi\varepsilon_r\varepsilon_0}\left[\frac{\partial\ln(\varepsilon_r)}{\partial p}\right]_T \frac{z_A^2 e^2}{a} \\ v^B &= v_*^B = \frac{4}{3}\pi N b^3 - \frac{N}{8\pi\varepsilon_r\varepsilon_0}\left[\frac{\partial\ln(\varepsilon_r)}{\partial p}\right]_T \frac{z_B^2 e^2}{b} \end{aligned} \tag{A24}$$

While introducing the intrinsic and Drude-Nernst contributions into the partial molar volume according to Eqs.(45), we arrive at Eq.(44) which is written for ion *A*, only.

*A.2.3 The high dilution limiting law for concentration dependency of apparent molar volume. Debye- Hückel approximation*

Now, we will determine the contribution into the density changes, $\delta d(\mathbf{r})$, due to the electric field produced by the Debye-Hückel screening space charge [79]. To this end, similarly to the above analysis, we consider hydrostatic equilibrium around an individual ion and complete the analyses based on Eq.(A16) by additional taking into account the electrical force acting the Debye-Hückel space charge which surrounds the ions and has the local density $\rho_{D-H}(\mathbf{r})$. Consequently, Eq.(A16) is modified, as

$$-\nabla p + \frac{1}{2}\nabla\left[E_{D-H}^2 \mathrm{d}\left(\frac{\partial \varepsilon_w}{\partial \mathrm{d}_w}\right)_T\right] - \frac{1}{2}E_{D-H}^2 \nabla \varepsilon_w + \mathbf{E}_{D-H}\rho_{D-H} = 0 \tag{A25}$$

In Eq. (A25),

$$\mathbf{E}_{D-H} = -\nabla \psi_{D-H} \tag{A26}$$

and, according to the Debye-Hückel theory [79],

$$\rho_{D-H} = -\varepsilon \kappa^2 \psi_{D-H}$$
$$\psi_{D-H} = \frac{ze}{4\pi\varepsilon_w(1+\kappa a)r}\exp\left[-\kappa(r-a)\right] \tag{A27}$$

In Eqs.(A27) $\kappa$ is the Debye-Hückel parameter [79]. For the binary electrolyte under consideration

$$\kappa^2 = \frac{F^2\left(z_A^2 c_A + z_B^2 c_B\right)}{\varepsilon_w RT} \tag{A28}$$

By combining Eqs.(25)-(28) we obtain

$$-\nabla\left(p - \frac{1}{2}\varepsilon_w\kappa^2\psi_{D-H}^2\right) + \frac{1}{2}\mathrm{d}\nabla\left[E_{D-H}^2\left(\frac{\partial \varepsilon_w}{\partial \mathrm{d}_w}\right)_T\right] = 0 \tag{A29}$$

By timing both the sides of Eq. (A29) by $(\partial \mathrm{d}/\partial p)_T$ and ignoring the volumes of screening ions, i.e., assuming that $\mathrm{d} = \mathrm{d}_w(p)$, one obtains

$$-\nabla \mathrm{d}_w + \frac{1}{2}\left(\frac{\partial \mathrm{d}_w}{\partial p}\right)_T \varepsilon\kappa^2\nabla\psi_{D-H}^2 + \frac{1}{2}\left(\frac{\partial \mathrm{d}_w}{\partial p}\right)_T \mathrm{d}\nabla\left[E_{D-H}^2\left(\frac{\partial \varepsilon_w}{\partial \mathrm{d}_w}\right)_T\right] = 0 \tag{A30}$$

As $\delta\varepsilon(r)/\varepsilon_w == O(E^2 \partial\varepsilon/\partial p)$ and $\delta \mathrm{d}/\mathrm{d}_w = O(E^2 \partial\varepsilon/\partial p)$, we retain the first order perturbation terms, only, to obtain

$$-\nabla\left[\delta d - \frac{1}{2}\left(\frac{\partial d_w}{\partial p}\right)_T \varepsilon\kappa^2\psi_{D-H}^2 - \frac{1}{2}d_w E_{D-H}^2\left(\frac{\partial \varepsilon_w}{\partial p}\right)_T\right] = 0 \qquad (A31)$$

By integrating Eq.(A31), we obtain

$$\frac{\delta d(\mathbf{r})}{d_w} = \frac{\delta d_{N-D}(\mathbf{r})}{d_w} + \frac{\delta d_{R-R}(\mathbf{r})}{d_w} \qquad (A32)$$

where $\delta d_{D-N}(\mathbf{r})$ and $\delta d_{R-R}(\mathbf{r})$ are the local perturbations of density predicted from the Drude-Nernst and Redlich-Rosenfeld models, respectively. The first one, $\delta d_{D-N}(\mathbf{r})$ is given by Eq.(A21), and the second one is represented with the help of Eqs.(A21), (A31) and (A32), as

$$\frac{\delta d_{R-R}(\mathbf{r})}{d_w} = \frac{1}{2}\beta\varepsilon\kappa^2\psi_{D-H}^2(\mathbf{r}) + \frac{1}{2}\left(\frac{\partial \varepsilon_w}{\partial p}\right)_T\left[E_{D-H}^2(\mathbf{r}) - E^2(\mathbf{r})\right] \qquad (A33)$$

where $E(\mathbf{r})$ is the Columbic field strength magnitude given by Eq.(A19) and $\beta = (\partial d_w/\partial p)_T/d_w$ is the solvent compressibility.

By combining Eqs.(A15), (A32) and (A33), we obtain for the reorganization part apparent molar volume

$$v_{reoorq}^A = v_A(0) - N\beta\frac{(ze)^2}{16\pi\varepsilon_w(1+\kappa a)^2}\kappa + \\ 4\pi N\frac{1}{2}\left(\frac{\partial \varepsilon_w}{\partial p}\right)_T\left[\frac{ze}{4\pi\varepsilon_w(1+\kappa a)}\right]^2 \int_a^\infty \frac{(1+\kappa a)^2 - (1+\kappa r)^2 \exp[-2\kappa(r-a)]}{r^4}r^2 dr \qquad (A34)$$

Passing to the limit $\kappa a \to 0$,

$$v_{reoorq}^A = v_A(0) + N\frac{3(z_A e)^2}{16\pi\varepsilon_w}\kappa\left[\frac{1}{\varepsilon_w}\left(\frac{\partial \varepsilon_w}{\partial p}\right)_T - \frac{\beta}{3}\right] \\ v_{reoorq}^B = v_B(0) + N\frac{3(z_B e)^2}{16\pi\varepsilon_w}\kappa\left[\frac{1}{\varepsilon_w}\left(\frac{\partial \varepsilon_w}{\partial p}\right)_T - \frac{\beta}{3}\right] \qquad (A35)$$

Consequently, by using Eqs.(5), (A24), (A28), (A35) and recalling that $\beta = (\partial d_w/\partial p)_T/d_w$, one obtains

$$v = v(0) + \left(\frac{\nu_A z_A^2 + \nu_B z_B^2}{4\varepsilon_w}\right)^{3/2} \frac{F^3}{N\pi\sqrt{RT}}\left[\frac{1}{\varepsilon_w}\left(\frac{\partial \varepsilon_w}{\partial p}\right)_T - \frac{1}{3d_w}\left(\frac{\partial d_w}{\partial p}\right)_T\right]\sqrt{c} \qquad (A36)$$

A comparison of Eqs.(A36) and (27) leads to Eq.(46).

**References**

1. Lakshminarayanaiah N: Transport Phenomena in Membranes. Academic Press; 1969.
2. Strathmann H : Ion-Exchange Membrane Separation Processes. Elsevier; 2004.

3. Mulder M: Basic Principles of Membrane Technology. Kluwer Academic Publishers; 1996.
4. Morf WE: The Principles of Ion-Selective Electrodes and of Membrane Transport, Studies in Analyical Chemistry, v. 2. Elsevier Science; 1981.
5. Rudin A, Choi P: The Elements of Polymer Science and Engineering. Oxyford: Academic Press; 2013.
6. Pattle RE, Nature, 1954; 174: 660. https://doi.org/10.1038/174660a0
7. Norman RS, Science, 1974; 186: 350. https://doi.org/10.1126/science.186.4161.350
8. Loeb S, Science, 1975; 189: 654. https://doi.org/10.1126/science.189.4203.654
9. Długołecki P, Nymeijer K, Metz S, Wessling M, Journal of membrane Science, 2008; 319: 214. https://doi.org/10.1016/j.memsci.2008.03.037
10. Veerman J, Saakes M, Metz SJ, Harmsen GJ, Journal of membrane Science, 2009; 327: 136. https://doi.org/10.1016/j.memsci.2008.11.015
11. Logan BE, Elimelech M, Nature, 2012; 488: 313. https://doi.org/10.1038/nature11477
12. Vermaas DA, Guler E, Saakes M, Nijmeijer K, Energy Procedia, 2012; 20: 170. https://doi.org/10.1016/j.egypro.2012.03.018
13. Yip NY, Vermaas DA, Nijmeijer K, Elimelech M, Environ. Sci. Technol., 2014; 48: 4925. https://doi.org/10.1021/es5005413
14. Yip NY, Elimelech M, Environ. Sci. Technol., 2014; 48: 11002. https://doi.org/10.1021/es5029316
15. Zlotorowicz A, Strand RV, Burheim OS, Wilhelmsen Ø, Kjelstrup S, Journal of membrane Science, 2017; 523: 402. https://doi.org/10.1016/j.memsci.2016.10.003
16. Dukhin SS, Derjaguin BV, Electrokinetic phenomena. In Matijevic (editor), Surface and colloid science, vol. 7, Wiley; 1974.
17. Masliyah JH, Bhattacharjee S: Electrokinetic and colloid transport phenomena. John Wiley and Sons, Inc.; 2006.
18. Stillwel W: Introduction to Biological Membranes: Composition, Structure and Function. Elsevier Science; 2016.
19. de Groot SR, Masur P: Nonequilibrium Thermodynamics. Amsterdam: NorthHolland; 1962.
20. Prigogine I: Introduction to Thermodynamics of Irreversible Processes. Wiley Interscience; 1968.
21. Zholkovskij EK, Masliyah JH, Shilov VN and Bhattacharjee S, Advances in Colloid Interface Science, 2007; 134-135: 279 https://doi.org/10.1016/j.cis.2007.04.025
22. Filippov AN, Colloid Journal, 2018; 80: 716 https://doi.org/10.1134/S1061933X18060030
23. Onsager L, Phys. Rev., 1931; 37: 405. https://doi.org/10.1103/PhysRev.37.405
24. Onsager L, Phys. Rev., 1931; 38: 2265. https://doi.org/10.1103/PhysRev.38.2265
25. Mazur P, Overbeek JThG, Rec. Trav. Chim., 1951; 70: 83. https://doi.org/10.1002/recl.19510700114
26. Lorenz P, J. Phys. Chem., 1952; 56: 775. https://doi.org/10.1021/j150498a031
27. Staverman AJ, Trans. Faraday Soc., I952; 48: 176. https://doi.org/10.1039/TF9524800176
28. Kedem O, Katchalsky A, Biochimica et Biophysica Acta, 1958; 27: 229. https://doi.org/10.1016/0006-3002(58)90330-5
29. Katchalsky A, Curran PF: Non-equilibrium Thermodynamics in Biophysics. Harvard University Press; 1965.
30. Michaeli I, Kedem O, Trans. Faraday Soc., 1961; 57: 1185. https://doi.org/10.1039/TF9615701185
31. Kedem O, Katchalsky A, Trans. Faraday Soc. , 1963; 59: 1918. https://doi.org/10.1039/TF9635901918

32. Weinstein JN , Caplan SR, J. Phys. Chem., 1973; 77: 2710. https://doi.org/10.1021/j100640a030
33. Kedem O, Comments. J Phys. Chem., 1973; 77: 2711. https://doi.org/10.1021/j100640a031
34. Caplan SR , Essig A: Bioenergetics and Nonequilibrium Thermodynamics. Harvard University Press; 1983.
35. Yaroshchuk AE, Adv. Colloid Interface Sci.,1995; 60: 1. https://doi.org/10.1016/0001-8686(95)00246-M
36. Yaroshchuk AE, J Membr Sci, 2002; 198: 285. https://doi.org/10.1016/S0376-7388(01)00668-8.
37. Nikonenko V, Zabolotsky V, Larchet C, Auclair B, Pourcelly G, Desalination, 2002; 147: 369. https://doi.org/10.1016/S0011-9164(02)00611-2
38. Auclair B, Nikonenko V, Larchet C, Métayer M, and Dammak L, Journal of membrane Science, 2002; 195: 89. https://doi.org/10.1016/S0376-7388(01)00556-7
39. Larchet C, Auclair B, Nikonenko V, Electrochimica Acta, 2004; 49: 1711. https://doi.org/10.1016/j.electacta.2003.11.030
40. Koter S, Desalination, 2006; 198: 335. https://doi.org/10.1016/j.desal.2006.02.009
41. Koter S, Kujawski W, Koter I, Journal of Membrane Science, 2007; 297: 226. https://doi.org/10.1016/j.memsci.2007.03.047
42. Yaroshchuk AE, Adv. Colloid Interface Sci, 2008; 139: 150. https://doi.org/10.1016/j.cis.2008.01.004
43. Bason S, Freger V, J Membr Sci, 2010; 360: 389. https://doi.org/10.1016/j.memsci.2010.05.037.
44. Fridman-Bishop N, Nir O, Lahav O, Freger V, Environ Sci Technol., 2015; 49: 8631. https://doi.org/10.1021/acs.est.5b00336.
45. Yaroshchuk AE, Bruening ML, Zholkovskiy E, Adv. Colloid Interface Sci, 2019; 268: 39. https://doi.org/10.1016/j.cis.2019.03.004
46. Kujawski W, Yaroshchuk A, Zholkovskiy E, Koter I, Koter S, Int. J. Mol. Sci., 2020; 21: 6325. https://doi.org/10.3390/ijms21176325
47. Fine R , Millero FJ, J. Chem. Phys, 1973; 59: 5529. https://doi.org/10.1063/1.1679903
48. Drude P, Nernst W, Zeitschrift für Physikalische Chemie, 1894; 15U: 79. doi:10.1515/zpch-1894-1506
49. Masson DO, The London, Edinburgh, and Dublin Philosophical Magazine and Journal of Science, 1929; 8: 218. https://doi.org/10.1080/14786440808564880
50. Redlich O, Rosenfeld P, Zeitschrift für Elektrochemie und angewandte physikalische Chemie, 1931; 37: 705. https://doi.org/10.1002/bbpc.19310370860
51. Redlich O, J. Phys. Chem., 1940; 44: 619. https://doi.org/10.1021/j150401a008
52. Owen BB, Brinkley SR, Chem. Rev., 1941; 29: 61, https://doi.org/10.1021/cr60094a003
53. Fajans K , Johnson O, Am. Chem. Soc. Jour.,1942; 64: 668 . https://doi.org/10.1021/ja01255a055
54. Owen BB, Brinkley SR, Ann. N.Y. Acad. Sci., 1949; 51: 753. https://doi.org/10.1111/j.1749-6632.1949.tb27303.x
55. Young TF, Smith MB, J. Phys. Chem, 1954; 58: 716. https://doi.org/10.1021/j150519a009
56. Couture AM, Laidler KJ, Canadian Journal of Chemistry, 1956; 34: 1209. https://doi.org/10.1139/v56-158
57. Mukerjee P, J. Phys. Chem., 1961; 65: 740. https://doi.org/10.1021/j100823a009
58. Redlich O, Meyer DM, Chem. Rev., 1964; 64: 221. https://doi.org/10.1021/cr60229a001
59. Glueckauf E, Trans. Faraday Soc., 1965; 61: 914. https://doi.org/10.1039/TF9656100914

60. Dunn LA, Trans. Faraday Soc., 1968; 64: 1898 https://doi.org/10.1039/TF9686401898
61. Cullen, PF: "Apparent molal volumes of some dilute aqueous rare earth salt solutions at 25°C " p, 110 (1969). *Retrospective Thesis and Dissertations. 4098.* https://lib.dr.iastate.edu/rtd/4098
62. Millero FJ, J. Phys. Chem., 1970; 74: 356. https://doi.org/10.1021/j100697a022
63. Millero FJ, Chem. Rev., 1971; 71: 147., https://doi.org/10.1021/cr60270a001
64. Millero FJ , Knox J, Journal of Chemical and Engineering Data,1973; 18: 407. https://doi.org/10.1021/je60059a023
65. Perron G, Desnoyers JE, Millero FJ, Canadian Journal of Chemistry,1974; 52: 3738. https://doi.org/10.1139/v74-558
66. Millero FJ, Surdo AL, Shin C, J. Phys. Chem.,1978; 82: 784. https://doi.org/10.1021/j100496a007
67. Koichiro M, Hiromitsu Y, Masayuki N, Bulletin of the Chemical Society of Japan, 1978 ; 51: 2508. https://doi.org/10.1246/bcsj.51.2508
68. Barlow RB, Br. J. Pharmac., 1980; 71: 17. https://doi.org/10.1111/j.1476-5381.1980.tb10904.x
69. Gonzalez-Perez A, Del Castillo J, Czapkiewicz J, Colloid and Polymer Science, 2002; 280: 503. https://doi.org/10.1007/s00396-001-0635-2
70. Klofutar C, Horvat J and Rudan-Tasič D, Acta Chim. Slov., 2006; 53: 274.
71. Marcus Y, J. Phys. Chem. B, 2009; 113: 10285. https://pubs.acs.org/doi/10.1021/jp9027244
72. Marcus Y, Chem. Rev, 2009; 109: 1346. https://doi.org/10.1021/cr8003828
73. Marcus Y, Chem. Rev., 2011; 111: 2761. https://doi.org/10.1021/cr100130d
74. Marcus Y: Ions in solution and their solvation. John Wiley & Sons, Inc.; 2015.
75. Kinart Z, Bald A, Physics and Chemistry of Liquids, 2011; 49: 366. https://DOI:10.1080/00319101003646579
76. Millero FJ: Chemical oceanography. Taylor & Francis Group, LLC; 2013. https://doi.org/10.1201/b14753
77. Millero FJ, J Solution Chem., 2014; 43: 1448. DOI 10.1007/s10953-014-0213-0
78. Zirman DU, Safarov JT, Dogani ÖM , Hassel EP, Uysal BZ, J. Serb. Chem. Soc., 2018; 83: 1005. https://doi.org/10.2298/JSC080517049U
79. Debye P, Hückel E, Physikalische Zeitschrift, 1923; 24: 185.
80. Nägele G, Physics Reports, 1996; 272: 215. https://doi.org/10.1016/0370-1573(95)00078-X
81. Landau LD, Lifshitz EM: Electrodynamics of continuous media. Pergamon; 1984.
82. Pauling L: The nature of the chemical bond and the structure of molecules and crystals; an introduction to modern structural chemistry. Cornell University Press; 1960. ISBN0-8014-0333-2.
83. Hittorf W: Poggendorf Annalen, (Annalen der Physik und Chimie), 1853; 89: 177.
84. Hittorf W: Poggendorf Annalen, (Annalen der Physik und Chimie), 1856; 98: 1.